\documentclass[runningheads]{llncs}

\usepackage[T1, T2A]{fontenc}
\usepackage[utf8]{inputenc}
\usepackage{graphicx}
\usepackage{booktabs}
\usepackage{amsmath}
\usepackage{xcolor}
\usepackage{multirow}
\usepackage{diagbox}
\usepackage{dirtytalk}
\usepackage{academicons}
\usepackage{colortbl}
\usepackage[most]{tcolorbox}
\usepackage{enumitem}
\usepackage{xurl}
\usepackage{cite}
\usepackage{csquotes}
\usepackage{listings}
\usepackage[title]{appendix}
\usepackage{tablefootnote}
\usepackage{algorithm}
\usepackage{algpseudocode}
\usepackage{epigraph}
\usepackage{booktabs}
\usepackage{subfig}
\usepackage{shapepar}
\usepackage{float}
\usepackage[depth=3]{bookmark}
\usepackage{minitoc}
\usepackage{pifont}
\usepackage{orcidlink}
\usepackage{bm}
\usepackage{tikz}
\usepackage{tikz-imagelabels}
\usepackage[table,xcdraw]{xcolor}
\usepackage{longtable}
\usepackage{dingbat}
\usepackage[hyperpageref]{backref}
\usepackage[normalem]{ulem}

\definecolor{adversarial}{HTML}{971a1e}
\definecolor{color_a}{HTML}{893244}
\definecolor{color_b}{HTML}{4F404C}
\definecolor{color_c}{HTML}{007F7F}
\definecolor{color_d}{HTML}{6082B6}
\definecolor{color_e}{HTML}{819171}
\definecolor{color_f}{HTML}{867d99}
\definecolor{color_g}{HTML}{28465B}

\urldef{\translateLocally}\url{https://github.com/XapaJIaMnu/translateLocally}
\urldef{\Ollama}\url{https://ollama.com/}
\urldef{\LLMModel}\url{https://huggingface.co/Jackrong/Negentropy-claude-opus-4.7-9B-GGUF}
\urldef{\PEGASUS}\url{https://huggingface.co/docs/transformers/model_doc/pegasus}
\urldef{\pyUnicodeSteganography}\url{https://github.com/bunnylab/pyUnicodeSteganography}
\urldef{\SilverSpeak}\url{https://github.com/ACMCMC/silverspeak}
\urldef{\eng}\url{https://github.com/orsinium-labs/eng/}
\urldef{\MullvadVPN}\url{https://mullvad.net/en/blog/mullvad-vpn-was-subject-to-a-search-warrant-customer-data-not-compromised}
\urldef{\CoverYourTracks}\url{https://coveryourtracks.eff.org/}
\urldef{\Tor}\url{https://support.torproject.org/about-tor/how-tor-works/overview/}
\urldef{\MaliciousTorRelays}\url{https://blog.torproject.org/malicious-relays-health-tor-network/}
\urldef{\PGPTutorial}\url{https://emailselfdefense.fsf.org/en/}
\urldef{\Signal}\url{https://signal.org/download/}
\urldef{\GrapheneOS}\url{https://grapheneos.org/install/}
\urldef{\LinuxDistributions}\url{https://distrowatch.com/dwres.php?resource=beginners-why}
\urldef{\EmailAliasing}\url{https://www.privacyguides.org/en/email-aliasing/}
\urldef{\DNSFiltering}\url{https://www.privacyguides.org/en/advanced/dns-overview/}
\urldef{\DNS}\url{https://www.privacyguides.org/en/dns/}
\urldef{\ContactRepresentative}\url{https://www.eff.org/congress/}
\urldef{\VPSVPN}\url{https://www.bluehost.com/blog/vps-for-vpn/}
\urldef{\Browser}\url{https://www.privacyguides.org/en/desktop-browsers/}
\urldef{\CanaryToken}\url{https://docs.canarytokens.org/guide/}
\urldef{\VPN}\url{https://ssd.eff.org/module/choosing-vpn-thats-right-you}

\renewcommand*{\backref}[1]{}
\renewcommand*{\backrefalt}[4]{%
    \ifcase #1%
        \or Cited on Page~#2.%
        \else Cited on Pages~#2.%
    \fi%
}

\title{
    Occluded Oculus: Operationalizing Stylistic Obscurement
}

\author{
    Robert Dilworth \orcidlink{0009-0005-5497-9810}
}

\authorrunning{
    Robert Dilworth
}
\titlerunning{
    Occluded Oculus
}

\institute{
    Department of Computer Science and Engineering, Mississippi State University, Mississippi State, Mississippi, USA\\
    \email{rkd103@msstate.edu}
}

\hypersetup{
    pdftitle={Occluded Oculus: Operationalizing Stylistic Obscurement},
    pdfsubject={cs.CR, cs.CL, cs.IR},
    pdfauthor={Robert Dilworth},
    pdfkeywords={Privacy, Adversarial Stylometry, Stylometry, Steganography, Zero-width Unicode Characters, Homoglyph Substitution, Translation, Obfuscation, Imitation, Injection, Authorship Verification, Authorship Attribution, TraceTarnish},
    colorlinks=true,
    linkcolor=color_a,
    citecolor=color_c,    
    urlcolor=color_d,
}

\begin{document}

\maketitle

\begin{abstract}

    What did it take for Hermes, the devout messenger of the Olympian gods, to slay Argus Panoptes, the multi-eyed giant of Greek myth\footnote{The mythological account was sourced from \textit{Mellenthin and Shapiro} \cite{Mellenthin2017}, but, as with such works, numerous similar yet distinctly different retellings of the events exist.}? As the perfect guardian, Panoptes' legion of ever-watchful eyes proved difficult---but not impossible---to defeat. The centerpiece of Hermes' strategy was obfuscation and sabotage. Posing as a shepherd, Hermes sealed each of Panoptes' eyes---eyes that would otherwise have alerted the fearsome giant to Hermes' plot---and vanquished him. The moral of the story: when a challenger must surmount a formidable foe---one far greater in stature and vastly more equipped---crafty maneuvers are not merely advisable but indispensable for victory. In this work, the ``challenger'' is a collective leveraging adversarial tactics to overcome the ``multi-eyed giant'' of stylometric systems and surveillance apparatuses. To successfully claw back the privacy siphoned by the multi-eyed giant, the challenger must carefully evaluate their plan of attack, \textsc{TraceTarnish}, and determine what does and does not work to anonymize the authorship of text. To that end, we conduct an ablation study of \textsc{TraceTarnish} to better understand which module---Translation, Obfuscation, Imitation, or Injection---best confounds a stylometric system. Our results indicate that the most effective approach was Injection, meaning that inserting zero-width Unicode characters, homoglyphs, and intentional misspellings neutralizes the indefatigable eyes long enough to claim the head of the all-seeing giant.

    
\end{abstract}

\section{Ablating the Components of Our \textsc{TraceTarnish} Attack to Gauge Granular Adversarial Effect}
\label{sec:Ablating_Components}

    \epigraph{\textcolor{adversarial}{Nothing is absolute. Everything changes, everything moves, everything revolves, everything flies and goes away.}}{\textit{{\scriptsize Frida Kahlo}}} 
    
    Authorship attribution \textit{is} a supervised text-analysis task that determines the most likely author of an anonymous or disputed document by comparing its stylometric features---such as function-word frequencies, character- and word-level \( n \)-grams, part-of-speech patterns, and lexical richness---to a corpus of texts with known authors, using statistical or machine-learning classifiers to assign the document to the author whose writing style it most closely matches (\textit{Savoy} \cite{Savoy2020}).

    Authorship attribution \textit{is not}, however, a fixed, immutable fact. It depends on the linguistic features and statistical models employed, the amount and type of text available, and the similarity among potential authors. Because these factors can change---new texts may be added, models can be refined, and subtle stylistic variations can be introduced---the confidence in \textit{any} attribution can vary over time. Thus, the certainty of an attribution is \textit{never} absolute and is, at best, fluid rather than definitive.
    
    Aiming to target that fluidity, we continue work on \textsc{TraceTarnish}---our adversarial stylometry\footnote{For definitions of the relevant terms used throughout this work, refer to (\textbf{Appendix \ref{appx:Definitions}}).} attack (\textit{Dilworth} \cite{UnicodeAdversarialStylometry2025,TraceTarnish2025,StegoStylo2026,HijackingTextHeritage2026}).

    By subtly perturbing stylistic cues and embedding mutable artifacts, the attack strives to make the results of stylometric analysis inconsistent whilst secondarily polluting textual data, nudging attribution toward uncertainty and affording plausible deniability without overtly breaking the text. Like most security measures, its primary purpose is to add \textit{friction}: it seeks to make attribution more difficult, though not impossible, for a would-be adversary, thereby raising the statistical threshold the adversary must overcome. If sufficient plausible deniability is generated by performing the attack---enough to enable repudiation---then the attack can be deemed successful, as any failure of identification, in any capacity, should be interpreted as a win.

    This does not preclude the myriad ways in which our proposed attack could be dismantled or rendered ineffective through data hygiene, including but not limited to: incorporating Unicode-aware preprocessing in favor of standard normalization, employing robust word-segmentation and cleaning pipelines, running character-level feature extraction to discard non-canonical glyphs, and enforcing platform-level sanitation to trivially strip (or prohibit) zero-width characters and homoglyphs. Any one of these measures could neutralize the Injection component of our attack.

    That being said, we stress that the attack's goal is to \textit{reduce} attribution accuracy, and it is only practically feasible when platforms accept multilingual text---raising the likelihood that homoglyphs survive---and when zero-width characters are retained for legitimate purposes such as right-to-left scripts, a situation increasingly common as services expand support for diverse languages to provide a seamless user experience.

    The attack becomes practical when those two conditions hold, and with most platforms vying for larger user bases, they are incentivized to ensure an unencumbered environment for all patrons, which necessarily means multi-language support, introducing the leeway that our attack---or at least the Injection component---can exploit.
    
    A reasonable assumption we hope to capitalize on is \textit{effort}: given the opportunity and depending on the recipient's disposition and the platform facilitating the exchange, most people who \textit{could be} involved will likely take shortcuts whenever possible. If minimal effort is applied to scrutinizing our attack's outputs, the chance of success raises. A more vigilant, skeptical, and distrustful recipient or platform would be better poised to neutralize the attack, given advanced warning and adequate precautions.

    In any event, the next stage of our work will be an ablation study, in which the four \textsc{TraceTarnish} components---Translation, Imitation, Obfuscation, and Injection---are evaluated. The marginal contribution of Injection relative to the remaining three components, and vice versa for the remaining techniques, will be directly measured and reported, constituting the main contribution of this work.

    \begin{figure}[H]
        \centering
        \includegraphics[width=0.87\linewidth]{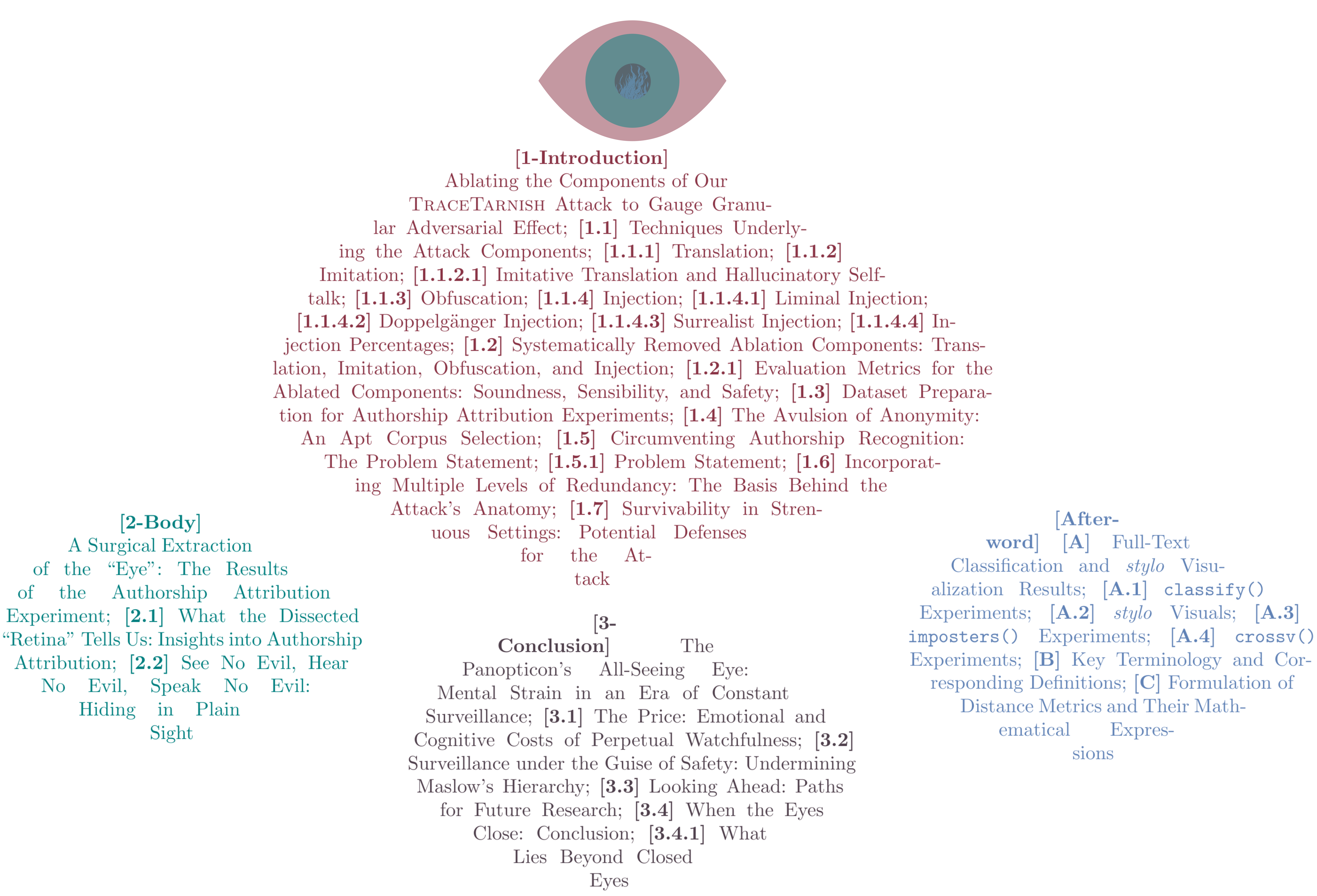}
        \caption{A stylized visual illustrating the paper's logical framework and organizational structure.}
        \label{fig:Paper_Structure}
    \end{figure}

    \subsection{Techniques Underlying the Attack Components}
    \label{subsec:Technique_Components}

        Before we go further, we first cleave our attack into its constituent components. Each element's function and the tools used to achieve the desired adversarial effect are described to help contextualize their roles in the overall attack.

        \subsubsection{Translation.}
        \label{subsubsec:Translation}

            We used \textit{translateLocally}\footnote{\translateLocally} to sequentially chain translations to and from multiple languages. The round-trip sequence used throughout the study was English \( \rightarrow \) Spanish \( \rightarrow \) German \( \rightarrow \) English. We chose an open-source solution over household names to reduce data exposure; locally executing personally vetted, freely available code is far superior to relying on opaque boxes that may leak data.

        \subsubsection{Imitation.}
        \label{subsubsec:Imitation}

\begin{figure}[H]
                \centering
                \includegraphics[width=0.6952\linewidth]{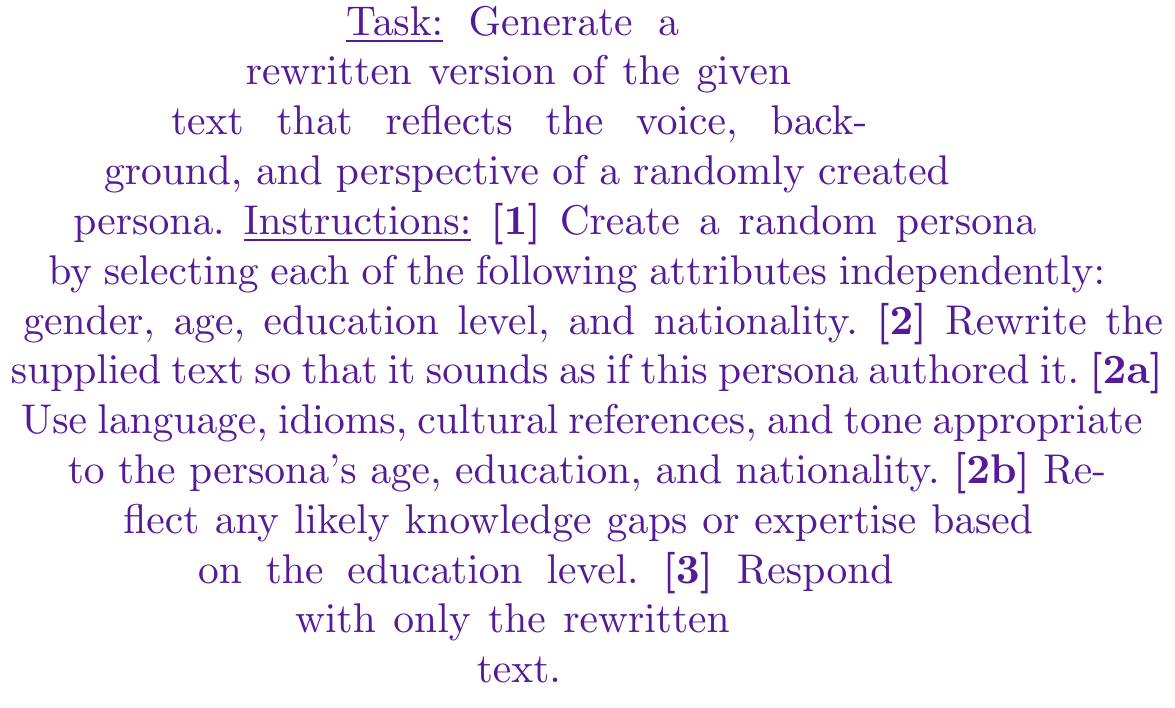}
                \caption{The prompt issued to our offline, self-hosted LLM that facilitates the adversarial technique of Imitation. A \textit{helter-skelter} selection of persona is desirable, provided it creates a stylometric distance from the original text fed to the LLM.}
                \label{fig:Imitation_LLM_Prompt}
            \end{figure}

\begin{figure}[H]
                \centering
                \includegraphics[width=1\linewidth]{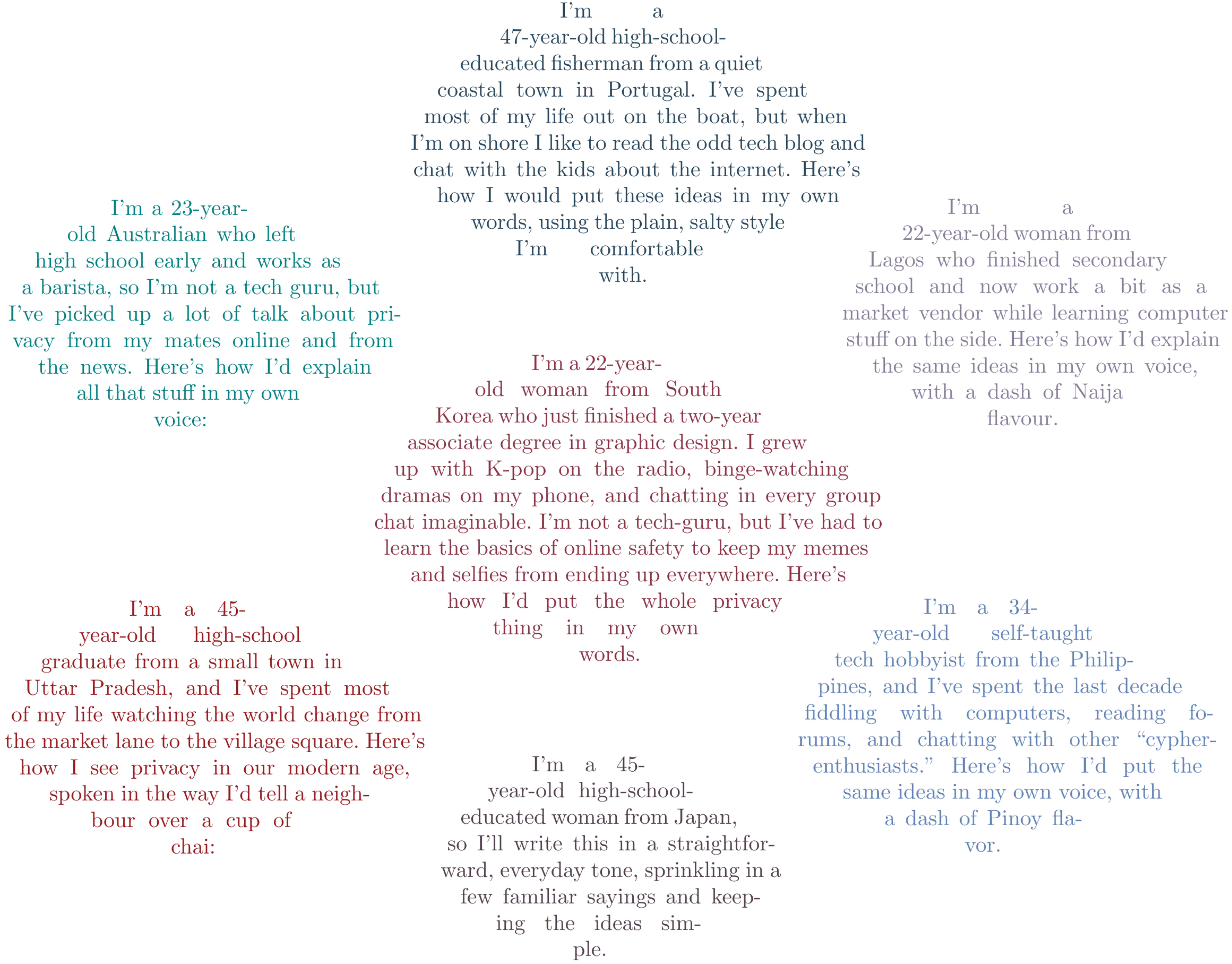}
                \caption{A showcase\protect\footnotemark of the LLM's ``persona preambles.'' Upon reflection, there appears to be a bias in the selection of ``nationality,'' with the LLM choosing a Brazilian persona more often than any other option. In most runs, the initial persona the model generates belongs to that demographic. For this iteration, the model assumed \textcolor{color_c}{Australian}, \textcolor{color_g}{Portuguese}, \textcolor{color_f}{Nigerian}, \textcolor{color_d}{Filipino}, \textcolor{color_b}{Japanese}, \textcolor{adversarial}{Indian}, and \textcolor{color_a}{South Korean} personas. Perhaps this observation is worth delving into. More importantly, it is worth acknowledging that the caricatures of the various personas and the stereotypes depicted in the model's responses reflect the model, not the authors. As outsiders, we are oblivious to them and make no claims about their validity.}
                \label{fig:Persona_Preambles}
            \end{figure}

            We employed a self-hosted, locally run Large Language Model (LLM) via Ollama\footnote{\Ollama} to achieve Imitation, transforming text so that it reads as if written by another entity. The model used throughout was \textit{Negentropy-claude-opus-4.7-9B-GGUF}\footnote{\LLMModel}. Our hardware and computational capabilities strongly influenced the model selection; the trade-off between speed and output quality, as well as CPU consumption, guided our choice. Using Ollama instead of a direct cloud-based offering helped avoid non-consensual data training, ensuring data minimization and privacy. The same prompt---see (\textbf{Figure \ref{fig:Imitation_LLM_Prompt}})---was used to achieve the desired effect, and all ``persona preambles'' were removed from the final responses (see (\textbf{Figure \ref{fig:Persona_Preambles}})). Rather, our prompt asks the LLM to assume a random personality, and we observed that it often prefaced its response with the persona's backstory.

            \footnotetext{The composition and layout of the figure was loosely influenced by Dadaist artist \href{https://www.artchive.com/artwork/strauss-hannah-hoch-1965/}{Hannah H\"och}.}

            \paragraph{Imitative Translation and Hallucinatory Self-talk.}
            \label{para:Imitative_Translation}

                It is worthwhile to point out that the same reasoning behind round-trip translation could \textit{plausibly} be applied to our LLM-imitation approach. With Translation, the adversarial effect derives from the loss of meaning imparted by multiple rounds of machine translation. Imitation, in a way, could be structured similarly. Depending on the selection of persona (for clarification, the selection mechanism is not truly random), a forward-feeding stream of back-to-back LLM rewrites would \textit{probably} result in a stylometric equivalent of the ``telephone game'' (a game where participants pass a secret message by whispering it from one person to another, after which the initial wording is examined against the version that emerges at the end).

                When persona \( A \) whispers the text's contents to persona \( B \), persona \( A \)'s fabricated experiences would \textit{perhaps} influence how persona \( B \) interprets the message. The inaccurate but close-enough transference of information and its chained modifications from the personas would \textit{likely} lead to a similarly garbled result. 
                
                In this setup, the ``personas talking to themselves'' would be a tangible representation of prompting a self-hosted LLM with text whose authorship we wish to obscure: we take the output of the first exchange, feed it back into the model, and invoke a crafted prompt that instructs the LLM to adopt a new persona for each iteration. This recursive prompting would mirror the telephone-game dynamic while deliberately scrambling stylometric cues (\textit{Kandula et al.} \cite{Kandula2024}).

                Although, given that we are issuing these prompts on our own hardware, we will set the idea aside for the moment, as our compute is a prohibitive bottleneck.

                As we hope to demonstrate with a relatively lightweight, self-hosted model---and as \textit{Srivastava et al.} \cite{Srivastava2025} concluded from their research on then-frontier models---authorship impersonation via LLMs ``erodes the assumption that an authorial `fingerprint' is a stable and difficult-to-forge identifier.''

        \subsubsection{Obfuscation.}
        \label{subsubsec:Obfuscation}

            We used \textit{PEGASUS}\footnote{\PEGASUS} to programmatically paraphrase text line by line. An LLM could, admittedly, also fulfill this role if prompted appropriately (\textit{Fisher et al.} \cite{Fisher2024}). During review, we noticed that certain input sentences produced incoherent results, such as repetitions of the sequence ``888-353-1299'' in place of the expected summaries. We did not investigate the cause and chose to retain the questionable outputs.

        \subsubsection{Injection.}
        \label{subsubsec:Injection}

            We employed \textit{pyUnicodeSteganography}\footnote{\pyUnicodeSteganography} and \textit{SilverSpeak}\footnote{\SilverSpeak} to embed zero-width Unicode characters into the text (a Liminal Injection attack) and perform homoglyph substitutions (a Doppelg\"anger Injection attack), respectively. As a subtle form of misspelling (a Surrealist Injection attack), we applied \textit{eng}\footnote{\eng} to convert American English to British English in the texts.

            \textbf{Liminal Injection} is an adversarial technique that inserts zero-width Unicode characters (e.g., \texttt{[U+200F]}) as hidden infixes ``in-between'' the letters of words, creating a transitional layer that lies at the interstice between the visible text and the invisible characters---hence the name ``liminal.'' \textbf{Doppelg\"anger Injection}, drawing from its etymological namesake, performs homoglyph substitut\-ions---for example, replacing the Latin small letter ``o'' \texttt{[U+006F]} with the Cyrillic small letter ``о'' \texttt{[U+043E]}---a \textit{changeling-like} phenomenon that leaves behind a nearly indistinguishable substitute. \textbf{Surrealist Injection} is an adversarial technique that deliberately introduces unconventional misspellings into a text, echoing the whimsical defiance of the Surrealist painters---artistic renegades in their own right.
            
            \paragraph{Liminal Injection.}
            \label{para:Liminal_Injection}

                For further reading on text steganography, we recommend \textit{Ahvanooey et al.} \cite{Ahvanooey2019}. Their insights into zero-width Unicode characters and their observations of how platforms handled them were instrumental to our attack.

            \paragraph{Doppelg\"anger Injection.}
            \label{para:Doppelganger_Injection}

                While Doppelg\"anger Injection could be thwarted by converting the suspected text into an image and then using something akin to optical character recognition (OCR) to re-extract the text---detecting and recovering the letters that the homoglyphs resemble \cite{Woodbridge2018}---this adds further friction to the normalization process. Regarding the use of non-OCR normalization to perform the previously described cleansing, \textit{Cooper et al.} \cite{Cooper2023} remarked that ``no tool was found which automated the conversion of all possible homoglyphs into the Latin characters they resemble,'' after which a manual editorial effort was required to identify and remediate homoglyph-laden text. Scripting alone could only partially normalize the text.
                
                Although this technique could be used maliciously, infusing text with homoglyphs remains a proven strategy to avoid detection---and, by extension, to sabotage stylometric analysis.

            \paragraph{Surrealist Injection.}
            \label{para:Surrealist_Injection}

                Considering alternate spellings of words in the English lexicon as misspellings---as we have done in this study\footnote{By default, most spell-checkers treat alternate spellings as mistakes unless the user selects the corresponding regional dictionary.}---is due the deterministic nature of the machine tasked with analyzing the text. If it is looking to match all instances of ``hypothesize'' in a text and encounters ``hypothesise,'' there is potentially room for error in that discrepancy.

            \paragraph{Injection Percentages.}
            \label{para:Injection_Percentages}

                For clarity, our attack implements maximal levels of Injection, which means that almost every word in a text will contain zero-width Unicode characters, all other visible characters will be swapped with homoglyphs, and, where applicable, words with alternate spellings will be replaced (e.g., ``emphasise'' versus ``emphasize'').

    \subsection{Systematically Removed Ablation Components: Translation, Imitation, Obfuscation, and Injection}
    \label{subsec:Systematically_Removed}

\begin{figure}[H]
            \centering
            \includegraphics[width=1\linewidth]{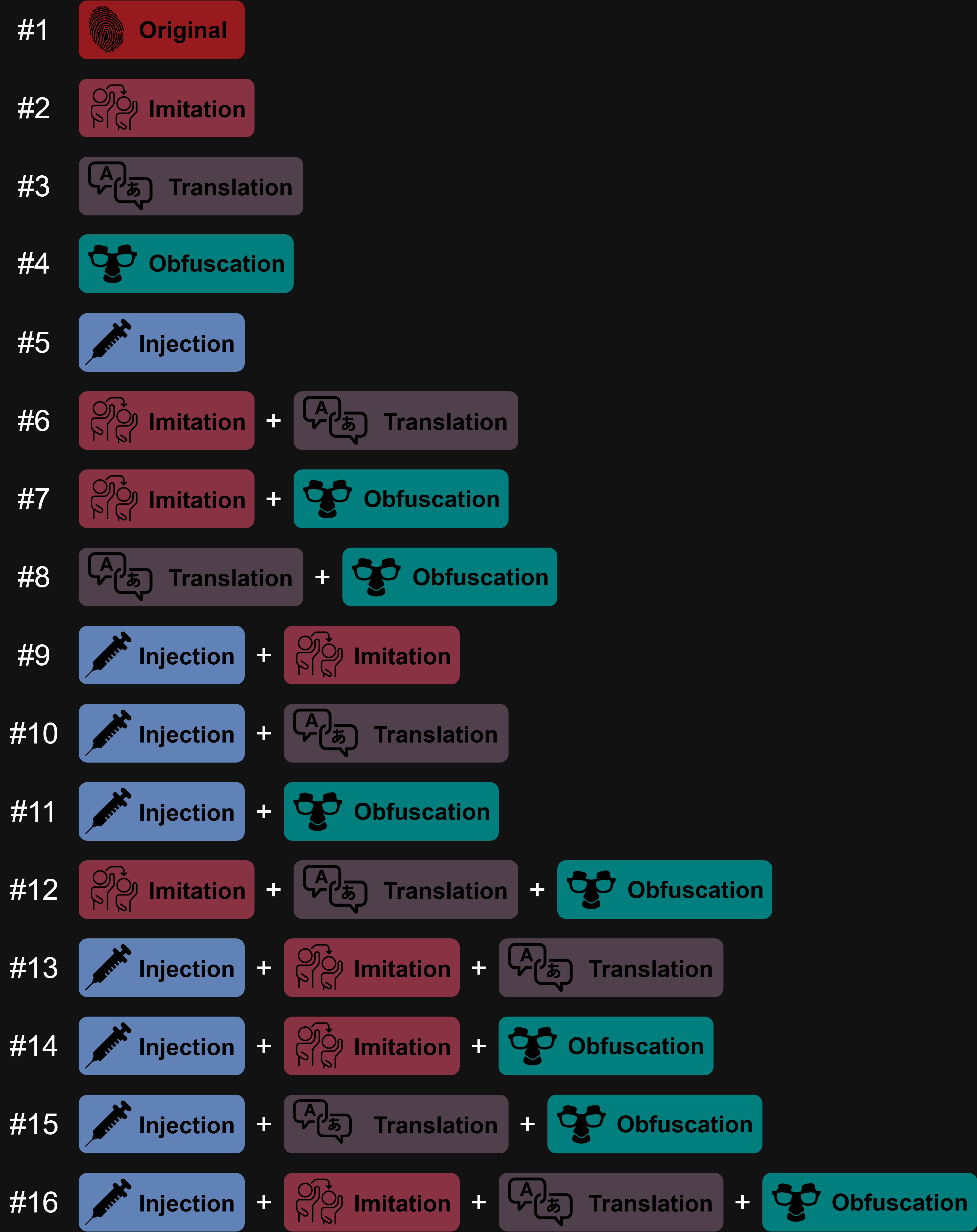}
            \caption{The components of our modular attack, \textsc{TraceTarnish}, are decomposed into distinct scenarios to quantify the adversarial effect achievable with the inclusion or exclusion of techniques.}
            \label{fig:Combinations}
            \vspace{1em}
        \end{figure}

        Here, we isolate the modular components comprising our \textsc{TraceTarnish} attack, as described in the preceding sections, to measure and compartmentalize their adversarial effect. Since we have broadly stratified the techniques into four clusters, we can derive fifteen scenarios, with the first being the control---the original, unaltered text. The cumulative integration of the four techniques---Translation, Imitation, Obfuscation, and Injection---constitutes a \textsc{TraceTarnish} attack. See (\textbf{Figure \ref{fig:Combinations}}).

        While the ordering in which each technique is applied is presumably significant, we will largely ignore this consideration. A clear example is any scenario involving Injection, which by its very nature and intended design, sabotages all other text-processing steps. In such cases, it makes sense for that step to take place last, as its potential to interfere with other steps would almost certainly produce undesirable results. In a similar vein, the same procedural treatment would likely also apply to Imitation, since this step was originally devised to smooth out the clunky, sometimes incomprehensible outputs produced by Translation and Obfuscation.

        \subsubsection{Evaluation Metrics for the Ablated Components: Soundness, Sensibility, and Safety.}
        \label{subsubsec:Evaluation_Metrics}
        
            To assess how each component affects the overall system, we evaluate them using three complementary metrics.

            \textbf{Soundness} ensures that the original meaning of texts, or their semantics, is not irreparably distorted by a technique. \textbf{Sensibility} ensures that metamorphosed texts remain human-readable and comprehensible. \textbf{Safety} ensures that texts are sufficiently resistant to de-obfuscation, encapsulating their ability to elude (re)attribution to their original sources of provenance.

            Translation and Obfuscation noticeably diminish a text's Sensibility, which correspondingly ushers in expected detriments to Soundness. Imitation, by its reliance on an LLM, tends to address those shortcomings, informing the decided-upon pipeline for \textcolor{color_e}{\textsc{TraceTarnish}}: \textcolor{color_b}{Translation} \( \rightarrow \) \textcolor{color_c}{Obfuscation} \( \rightarrow \) \textcolor{color_a}{Imitation} \( \rightarrow \) \textcolor{color_d}{Injection}.
            
            In terms of the previously specified tools, the settled-upon sequence of their use is: \textcolor{color_b}{\textit{translateLocally}} \( \rightarrow \) \textcolor{color_c}{\textit{PEGASUS}} \( \rightarrow \) \textcolor{color_a}{a self-hosted, offline LLM rewrite} \( \rightarrow \) \textcolor{color_d}{\textit{eng}} \( \rightarrow \) \textcolor{color_d}{\textit{pyUnicodeSteganography}} \( \rightarrow \) \textcolor{color_d}{\textit{SilverSpeak}}. It is paramount that the relied-upon LLM---though it need not be the one we selected---be fully under your control; likewise, the ordering of the Injection steps is relatively fixed. \textit{eng} will have trouble detecting words if they are poisoned beyond recognition by \textit{pyUnicodeSteganography} and \textit{SilverSpeak}, hence their sequencing.
            
            While Translation and Obfuscation positively impact Safety, that Safety comes at the expense of Soundness and Sensibility, which Imitation amends. From the perspective of a human observer, Injection has nearly zero impact on Soundness and Sensibility as we have described them (assuming proper rendering occurs), and it imparts a similar positive increase in Safety on account of the mechanics of Liminal and Doppelg\"anger Injection.

    \subsection{Dataset Preparation for Authorship Attribution Experiments}
    \label{subsec:Dataset_Preparation}

        We will get slightly ahead of ourselves here as we explain the collection of files prepared for the study. In the following section, we will go further in depth on why we chose the primary text and the additional texts used to facilitate the testing.

        As per \textit{stylo}'s documentation \cite{stylo2026}, the description of \texttt{classify()} is a ``function that performs a number of machine-learning methods for classification used in computational stylistics: [Burrows' Delta], k-Nearest Neighbors, Support Vector Machines, [Na\"{\i}ve] Bayes, and Nearest Shrunken Centroids.'' We provide this description verbatim to communicate that our results have been assessed by contemporary stylometric methods.

        In previous studies we relied upon the \texttt{imposters()} method, ``a machine-learning supervised classifier tailored to assess authorship verification tasks.'' The \texttt{imposters()} function was sufficient in those studies because the main goals were to establish the viability of the attack and demonstrate its confounding abilities. For those goals, an experimental setup focused on authorship verification was appropriate. Given multiple incrementally modified adversarial texts and the original, can we confirm that the adversarial texts still belong to the original author? The answer was \textit{no}, given sufficient Injection percentages (\textit{Dilworth} \cite{StegoStylo2026,HijackingTextHeritage2026}).

\begin{figure}[H]
            \centering
            \includegraphics[width=0.60\linewidth]{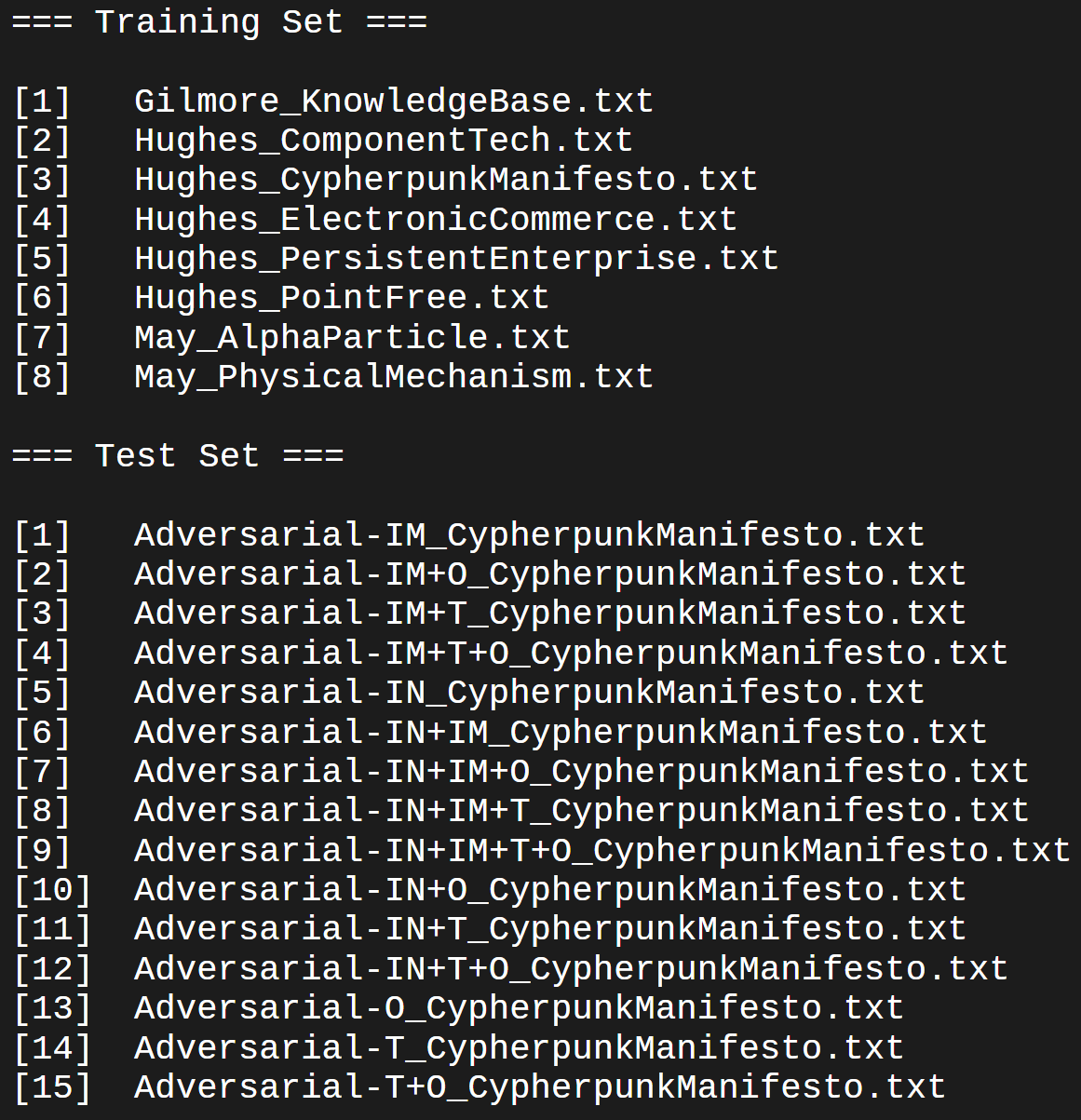}
            \caption{As \texttt{classify()} expects a training and test corpus, we have organized our files into suitably named directories, the contents of which are shown here. The training set consists of texts authored by John Gilmore \cite{Gilmore1984}, Eric Hughes \cite{Hughes1998,Hughes1999Persistent,Hughes1997,Hughes1999Point,Hughes1993}, and Timothy C. May \cite{May1979,May1978}. The test set comprises adversarially modified versions of Hughes's \textit{A Cypherpunk's Manifesto} \cite{Hughes1993}, which selectively and gradually applies the techniques that constitute our \textsc{TraceTarnish} attack, as visualized in (\textbf{Figure \ref{fig:Combinations}}).}
            \label{fig:Training_and_Test_Data_List}
        \end{figure}

\begin{figure}[H]
            \centering
            \includegraphics[width=0.60\linewidth]{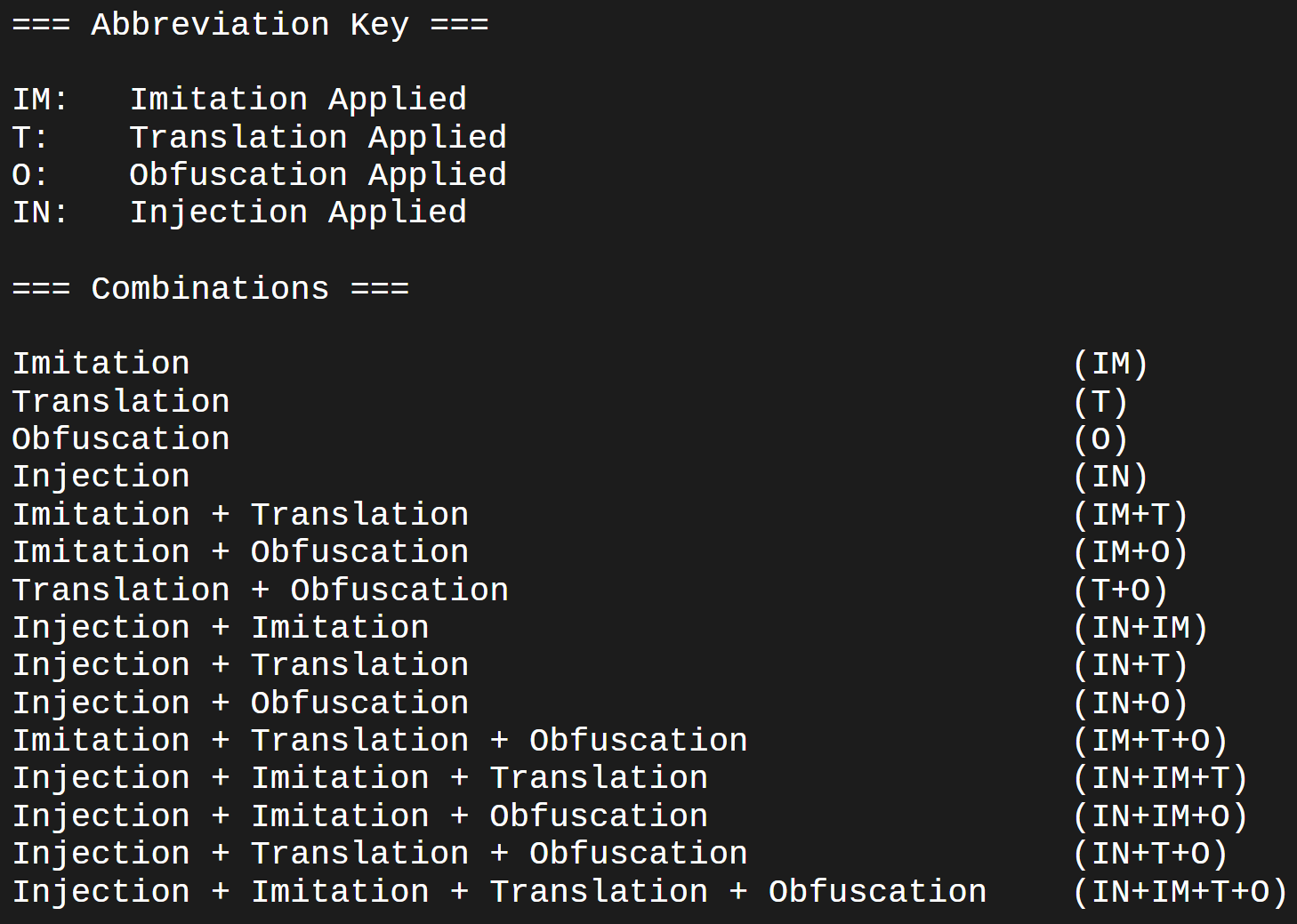}
            \caption{For ease of recognizing which files were associated with which scenario and which part of our attack was needed for their proper processing, we formulated a naming framework for the handled files.}
            \label{fig:File_Naming_Convention}
        \end{figure}

        Here, our task is authorship attribution. Given a corpus of texts authored by notable cypherpunk writers, can \textit{stylo}'s machine-learning classifier---\texttt{classify()}---accurately attribute the correct author to the adversarially modified texts, which have undergone differing adversarial treatments? To answer this question, we created a suitable corpus, as shown in (\textbf{Figure \ref{fig:Training_and_Test_Data_List}}). The adversarially treated files are presented in (\textbf{Figure \ref{fig:File_Naming_Convention}}).

        Concerning the file-naming convention and our use of \textit{stylo}, an R stylometry package \cite{Eder2016}, we adopt a schema that best interfaces with the package. The control files are prefixed with the author's surname, followed by an underscore and a condensed version of the text's title. All other files produced via adversarial tampering are prefixed with ``Adversarial.'' To add further descriptiveness to the filenames, we include abbreviations indicating the attack techniques applied to the text.

        We now circle back to that primary text and its appropriate selection for this study.

    \subsection{The Avulsion of Anonymity: An Apt Corpus Selection}
    \label{subsec:Avulsion_Anonymity}

        To conquer a fear---whatever it may be---it is often advised to run toward its source. Putting a face and a name to it, which essentially amounts to personification, is one of many steps to stymie its hold over an individual. Unless that recognition takes root, an issue remains locked in a mental vault---eternally relegated and unlabeled---so that the \textit{problem} does not become properly elevated to the status of a \textit{problem}.

        Everything to its proper place: identification comes before treatment. What is the problem? \textit{Pyrophobia}? What does that mean, and how does it affect you? Exposure therapy could help, but subjecting yourself to a ``trial by fire'' also has another prerequisite: a loss of privacy. Admitting the existence of the fear internally is a necessary step, but doing so externally would marginally improve the prognosis. When the internal acknowledgment stalls, the external reality soon catches up.

        It is precisely at this stage that we find ourselves. Without that acknowledgment, the issue continues to fester unnoticed, and the problem will persist indefinitely if it is not recognized as such. The attack on anonymity is one such problem.

        Talks of weakening and back-dooring encryption \cite{Klosowski2025}---a \textit{problem}. The outright banning or tightening of regulations on virtual private networks (VPNs) \cite{Alajaji2025}---a \textit{problem}. The broad enforcement of age and identity verification \cite{AV2026,Liu2026,Buckley2026,Mullin2026}---a \textit{problem}. The proliferation and indifference towards evolving surveillance capabilities \cite{Hamid2025,Wilkins2026,Maass2026,Brodkin2026,Cyphers2022,Ingraham2026,Goodin2015,Cox2026,Quintin2025,Liu2024,Guariglia2025,Koebler2026,Dupre2026,Pilkington2014,Feder2026,Koebler20260210,Adib2013,Guariglia2024}---a \textit{problem}. Our list could easily be expanded\footnote{The lesser of the ills---cognitive offloading enabled by AI---may be alleviated by a \textit{Hammurabian} response \cite{Goodin2026}, a response whose disproportionate---though perhaps justified---reciprocity is reminiscent of asphyxiation.} (adding LLMs capable of pinpointing authorship \cite{McArdle2026}---and traditional stylometry \cite{Emmery2024,Afroz2014,Nipa2023}---to the fray are no-brainers), but the point has been made: whatever \textit{illusion} we once had of a better future is being \textit{disillusioned} in near real time---a fleeting and dying \textit{will-o'-the-wisp}. However, to overcome \textit{something}, the \textit{thing} must first be recognized.

        While it does not address every issue we have raised, Eric Hughes's \textit{A Cypherpunk's Manifesto} serves as foundational kindling for that conversation. Even if it overlooks certain aspects---how could it not, given the time it was written?---it still offers a valuable touchstone. For this reason, we cyclically \textit{reduce}, \textit{reuse}, and \textit{recycle} the text as our corpus. The words shed light on a problem, and to suppress the problem we, too, must deprive it of oxygen.

        Stylometry, with its dualistic, \textit{agathokakological} nature, has the potential to become a problem: it can empower researchers to uncover hidden patterns and protect communities, yet it also furnishes adversaries with a precise map for tracking and profiling individuals.

        Armed with this awareness, we can direct our countermeasures. Our arsenal of adversarial techniques flares forward, forged specifically to \textit{snuff out} that fire---the exigent inferno that currently engulfs us with the \textit{Promethean} flame.

    \subsection{Circumventing Authorship Recognition: The Problem Statement}
    \label{subsec:Circumventing_Authorship}

        Having broadly introduced the study's \textit{core} problem, we now describe the \textit{specific} issue being examined.

        \subsubsection{Problem Statement.}
        \label{subsubsec:Problem_Statement}

            To communicate anonymously online in the public sph\-ere---without encryption\footnote{\PGPTutorial} or other cryptographic protocols\footnote{\Signal}---and under the assumption that anything public is, by definition, not private, which adversarial stylometry technique best serves that goal? Does a single technique suffice, or is a confluence of techniques used in tandem required? Which techniques are better suited for the task, and which are not? Is it feigning the idiosyncratic writing style of another entity (\textcolor{color_a}{Imitation})? Is it executing multiple rounds of machine translation from dissimilar languages (\textcolor{color_b}{Translation})? Is it paraphrasing the text to scrub the author's cadence and voice (\textcolor{color_c}{Obfuscation})? Or is it inserting invisible characters (zero-width Unicode characters), visually similar but differently interpreted characters (homoglyphs), or misspellings of words (\textcolor{color_d}{Injection})? In the face of such techniques, to what degree are the Soundness, Safety, and Sensibility of the text impacted?  

            For the sake of eliminating other externally identifying factors, we also assume that the user's IP address is masked by a reputable\footnote{\MullvadVPN} or self-hosted\footnote{\VPSVPN} VPN\footnote{\VPN}, a privacy-focused browser\footnote{\Browser} is used to connect to the platform facilitating the post---a browser that minimizes device fingerprinting\footnote{\CoverYourTracks} by reducing identification of operating system, user agent, time zone, etc.---Domain Name System-level filtering\footnote{\DNSFiltering} and encryption\footnote{\DNS} are in place, the operating system itself is hardened\footnote{\LinuxDistributions} and privacy-focused\footnote{\GrapheneOS}, the user adopts a pseudonym and aliased contact details (``mask'')\footnote{\EmailAliasing} to masquerade as someone else, and sufficient discipline is exercised to avoid cross-contaminating profiles and guises.

            The relevance of these measures is illuminated by \textit{Manish Tripathy} \cite{Tripathy2026}, which---though indirect---demonstrates why a public persona should \textit{never} share a digital fingerprint with a private persona.

    \subsection{Incorporating Multiple Levels of Redundancy: The Basis Behind the Attack's Anatomy}
    \label{subsec:Incorporating_Multiple}

        The problem statement introduces a facet worth highlighting.

        Envision this: your surroundings conceal a hidden menace---one that could rob you of your sight---and you are determined to safeguard your vision at all costs. A simple blindfold \textit{might} work, but it would obstruct your line of sight, which could, in turn, cause you to injure your eyes in unforeseen ways. Your goal: preserve your sight and avoid going blind\footnote{We recognize that our attack could have unintended consequences for valuable text-analysis applications such as screen readers. The potential collateral damage it may cause outweighs the primary motivation of privacy, as it could objectively worsen the experience of people with visual impairments and degrade engagement with social media and similar platforms \cite{Lee2022}. This trade-off highlights room for improvement; a better method would achieve the desired effect with pinpoint accuracy and avoid these drawbacks.}. Assessing your risk profile, you identify the immediate threats to your objective: ultraviolet radiation, particulate matter \cite{Ireland2026}, volatile chemicals, thermal exposure, intense light, infectious agents, mechanical trauma, ocular diseases, and \textit{arcane arts}.

        Beyond what can be achieved with regular examinations and proper nourishment, you settle upon four controls to manage the potential perils: eye drops, contact lenses, safety goggles, and face shields. Each, in isolation, offers only a modest degree of protection, but layering the safeguards proves prudent---the failure of any single control would not immediately expose you to the hazard(s).

        In practice, this is often expressed as \textbf{defense-in-depth} or a multi-layered control strategy. This approach builds redundancy and depth across various measures---technical, administrative, and physical. For our simplistic example, categorizing the controls by type of measure is beyond scope.

        With a plan of action formulated, you equip your armor. You apply eye drops to each eye, insert the contact lenses one by one, don your safety goggles, and slip on your face shield. In this state, a catastrophic system failure would have to occur to invalidate your suite of controls. Your eye drops' medicinal effect would need to expire, your contact lenses would need to become dislodged, your goggles would need to be ruptured, and your face shield would need to be pierced.

        A similar foundation fuels our attack's structure. If a platform or stylometric system purges zero-width characters, then there are the homoglyphs. If the homoglyphs are eradicated, then there are the misspellings. If the misspellings are accounted for, then there are the LLM rewrites. If the rewrites prove insufficient, then round-trip translation of the source text remains. If that too fails, then paraphrasing the source text ensures that at least some semblance of meaning is marred. In a scenario where the objective is to defeat stylometric analysis, this is what it means to incorporate defense-in-depth principles. The safety of your ``eyes'' depends on it, lest you risk losing your ``sight.''

    \subsection{Survivability in Strenuous Settings: Potential Defenses for the Attack}
    \label{subsec:Survivability}

    \textit{Risk}. Everything revolves around that ``globe,'' whose gravity attracts and repels \cite{McGruder2018} a host of dangers, both seen and unseen. Here we scrutinize the attack's weak points---weak points that can be exploited to cripple its malevolent effect.

    \begin{itemize}
        \item[\ding{118}] First, while homoglyphs slip past \textit{most} human readers, they are not invisible on \textit{every} platform. Various fonts render characters differently, and some environments expose Unicode explicitly. Thus, Doppelg\"anger Injection can become apparent to users of certain software or devices.
        \item[\ding{118}] Second, the attack targets primarily English-language detection systems. Many languages already employ multiple scripts, and their detectors tend to tolerate character variation. Thus, the generalizability of our findings to other languages remains uncertain.
    \end{itemize}

    Subsection \ref{subsec:Incorporating_Multiple} outlines the threat model in which the technique thrives; we have since highlighted where it may falter to encourage responsible disclosure. If a \textit{vulnerable party} deems the attack relevant to their threat model, transparent knowledge of its capabilities and limits may be what shields them from nightmarish horrors \cite{Prahlow2020}.

    A \textbf{threat model} is a systematic characterization of the adversarial landscape that delineates the capabilities, objectives, and constraints of potential attackers, as well as the assets and assumptions pertinent to the defended system. By explicitly enumerating \textit{who} might act maliciously, \textit{what} resources they can marshal, and \textit{which} vulnerabilities they might exploit, a threat model provides the analytical framework necessary to assess risk, prioritize defenses, and ensure the security measures align with the realistic dangers faced by the environment in question.

    As we mentioned in earlier sections, specialized preprocessing could trivially render the Injection component inert. \textit{Khan et al.} \cite{Khan2025}, however, present a series of pertinent counter-claims:

    ``[P]re-processing in stylometry can adversely impact the algorithms' performances.'' ``[S]pecial characters[---zero-width characters and homoglyphs---]can be vital in differentiating and recognizing an author's style.'' ``[O]ne author may use certain special characters [more frequently] compared to others.''

    Taken together, the implication here is that it is better to leave the text in its original state if the intention is to perform stylometric tasks such as authorship attribution, verification, or profiling. If the success of the analysis depends on processing raw text, and that raw text is ``dеаtһ-dealing'' \cite{Castagnaro2025} because it contains poison specifically designed to sabotage the analysis, then the prospects of our attack succeeding are far greater than originally anticipated. Keeping the text whole gives it the potential to cause damage; preprocessing the text to sanitize it reduces the efficacy of analysis.

    Just as gravity can both bind and release, a judicious mix of normalization and selective preservation can counteract the attack while retaining essential stylometric cues.

\section{A Surgical Extraction of the ``Eye'': The Results of the Authorship Attribution Experiment}
\label{sec:Surgical_Extraction}

    \epigraph{\textcolor{adversarial}{Is it possible that my brain, this precise, clean, glittering mechanism, like a chronometer without a speck of dust on it, is\ldots? Yes it is, now. I really feel there in the brain some foreign body like an eyelash in the eye. One does not feel one's whole body but this eye with a hair in it, one cannot forget it for a second\ldots}}{\textit{{\scriptsize We \\ Yevgeny Zamyatin}}}

    Of the many structures that make up the sensory organ called the eye, the retina is arguably the most important, as it contains the photoreceptor cells that convert light into electrical signals, which the brain, in turn, interprets as visual images. Without a functional retina, vision would be practically impossible---the likelihood of perceiving any coherent visual information drops to near zero.

    For our purposes, the structurally-related sentiment is transitive: if the ``eyes'' represent an author and the ``eye's perspective'' the stylistic signature we wish to blot out, then our adversarial undertaking could be likened to gouging out those eyes. Yes, they---by the association we have constructed---are indescribably irreplaceable, one of a kind. But the eyes should remain subject to the host, not the other way around. If they become a liability or antagonistic, plucking them out may be an option worth pursuing.

    The presumption here is that if your stylistic signature can be used against \textit{you}, then that risk supersedes any potential advantages it could have afforded \textit{you}. In such a case, if the risk cannot be transferred, reduced, or accepted---as would be the case in a low-risk landscape---a response would be necessary. Granted, we will refrain from illustrating such a high-risk scenario---\textit{or not}. Luckily for us, the motivating analogy of \textit{Alden Page} \cite{Page2015} more than suffices.

    From the perspective of the general populace, deliberately obfuscating your stylometric profile may come across as the ``paranoid machinations of the maniacal.'' For most, that assessment---while blunt---may hold some truth; it is markedly excessive, to the point of being overkill for most security scenarios. In almost every other circumstance, that level of concern is borderline irrational.

    However, that concern is not entirely misplaced or unwarranted. Want a curated experience? Share \textit{some} data. Want better recommendations? Share \textit{more} data. Want camaraderie? Share \textit{a deluge} of data that could later be hoovered up and used to identify you (if your earlier (in)voluntary revelations did not already satisfy that threshold).

    While more narrative scaffolding could potentially tie everything together, we will cease the analogy here and take a more direct approach.

    \subsection{What the Dissected ``Retina'' Tells Us: Insights into Authorship Attribution}
    \label{subsec:Dissected_Retina}

        For our experiment, we constructed a corpus of related authors and singled out a single text to undergo adversarial treatment with the intent of coercing \textit{stylo}'s \texttt{classify()} to misclassify its author (i.e., have the classifier reach an incorrect conclusion about the likely author). Besides populating the expected training and test-set parameters, we did not modify any other aspect of the function.

\begin{figure}[H]
            \centering
            \includegraphics[width=1\linewidth]{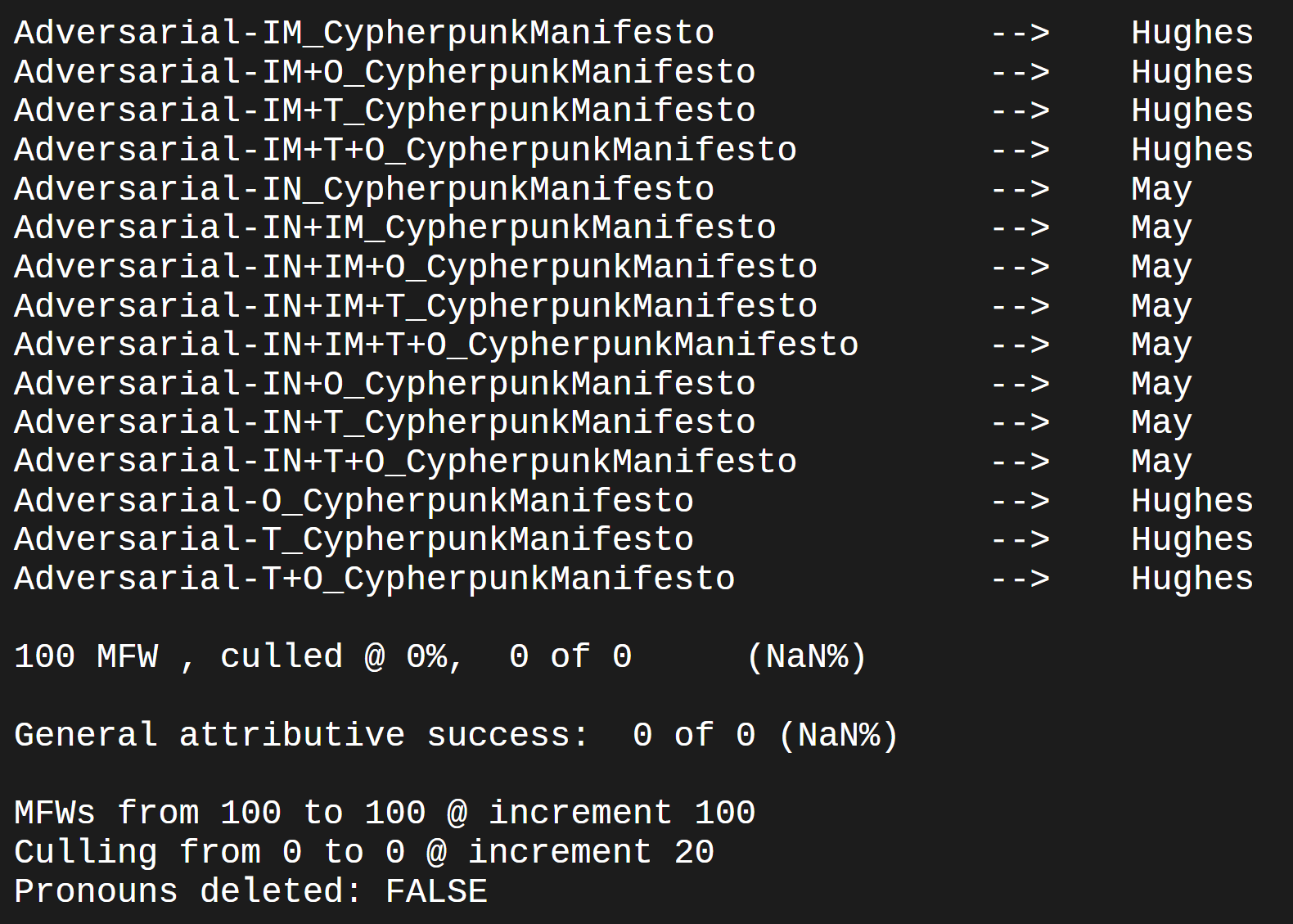}
            \caption{The final results of the \texttt{classify()} experiment show that only texts altered by Injection successfully induced misclassification.}
            \label{fig:Stylo_Classify_Final_Results}
        \end{figure}

        The outcome of the experiment roughly aligns with what could be anticipated from the setup and adversarial techniques employed. \textsc{TraceTarnish}'s additive adversarial effect overwhelmingly stems from the Injection component; it is the attack's essential component, as its sole application---without the other components of Translation, Obfuscation, and Imitation---resulted in misclassification (where misclassification means that May was attributed to Hughes's work). The final results of \texttt{classify()} corroborate this; see (\textbf{Figure \ref{fig:Stylo_Classify_Final_Results}}). (\textbf{Table \ref{tab:Stylo_Classify_Distance_Table}}) contains the computed distance table that influenced the attributions.

        \begin{table}[H]
            \resizebox{\textwidth}{!}{
                \begin{tabular}{|l|l|l|l|l|l|l|l|l|}
                    \hline
                    \rowcolor[HTML]{000000} 
                    {\color[HTML]{FFFFFF} } & {\color[HTML]{FFFFFF} \textbf{Gilmore}} & {\color[HTML]{FFFFFF} \textbf{Hughes}} & {\color[HTML]{FFFFFF} \textbf{Hughes}} & {\color[HTML]{FFFFFF} \textbf{Hughes}} & {\color[HTML]{FFFFFF} \textbf{Hughes}} & {\color[HTML]{FFFFFF} \textbf{Hughes}} & {\color[HTML]{FFFFFF} \textbf{May}} & {\color[HTML]{FFFFFF} \textbf{May}} \\ \hline
                    \cellcolor[HTML]{000000}{\color[HTML]{FFFFFF} \textbf{Adversarial-IM}} & 1.4809 & 1.3692 & 1.0477 & 1.3855 & 1.4421 & 1.3102 & 1.3495 & 1.3163 \\ \hline
                    \rowcolor[HTML]{867D99} 
                    \cellcolor[HTML]{000000}{\color[HTML]{FFFFFF} \textbf{Adversarial-IM+O}} & 1.6918 & 1.5171 & 0.9933 & 1.5046 & 1.6843 & 1.2794 & 1.5981 & 1.5649 \\ \hline
                    \cellcolor[HTML]{000000}{\color[HTML]{FFFFFF} \textbf{Adversarial-IM+T}} & 1.3875 & 1.2032 & 0.8119 & 1.1997 & 1.3188 & 1.0720 & 1.2391 & 1.2046 \\ \hline
                    \rowcolor[HTML]{867D99} 
                    \cellcolor[HTML]{000000}{\color[HTML]{FFFFFF} \textbf{Adversarial-IM+T+O}} & 1.4693 & 1.2625 & 0.9328 & 1.3659 & 1.4867 & 1.1840 & 1.3891 & 1.3546 \\ \hline
                    \cellcolor[HTML]{000000}{\color[HTML]{FFFFFF} \textbf{Adversarial-IN}} & 2.7552 & 3.0981 & 3.5869 & 3.5320 & 2.7861 & 3.4888 & 2.7051 & 2.6731 \\ \hline
                    \rowcolor[HTML]{867D99} 
                    \cellcolor[HTML]{000000}{\color[HTML]{FFFFFF} \textbf{Adversarial-IN+IM}} & 2.6720 & 3.0148 & 3.5036 & 3.4487 & 2.7028 & 3.4056 & 2.6218 & 2.5899 \\ \hline
                    \cellcolor[HTML]{000000}{\color[HTML]{FFFFFF} \textbf{Adversarial-IN+IM+O}} & 2.6568 & 2.9859 & 3.4747 & 3.4198 & 2.6877 & 3.3766 & 2.6066 & 2.5747 \\ \hline
                    \rowcolor[HTML]{867D99} 
                    \cellcolor[HTML]{000000}{\color[HTML]{FFFFFF} \textbf{Adversarial-IN+IM+T}} & 2.4820 & 2.8157 & 3.3045 & 3.2496 & 2.5129 & 3.2064 & 2.4318 & 2.3999 \\ \hline
                    \cellcolor[HTML]{000000}{\color[HTML]{FFFFFF} \textbf{Adversarial-IN+IM+T+O}} & 2.7325 & 3.0754 & 3.5642 & 3.5093 & 2.7634 & 3.4661 & 2.6823 & 2.6504 \\ \hline
                    \rowcolor[HTML]{867D99} 
                    \cellcolor[HTML]{000000}{\color[HTML]{FFFFFF} \textbf{Adversarial-IN+O}} & 2.8016 & 3.1445 & 3.6333 & 3.5784 & 2.8325 & 3.5352 & 2.7515 & 2.7195 \\ \hline
                    \cellcolor[HTML]{000000}{\color[HTML]{FFFFFF} \textbf{Adversarial-IN+T}} & \textbf{2.8497} & \textbf{3.1926} & \textbf{3.6814} & \textbf{3.6265} & \textbf{2.8806} & \textbf{3.5833} & \textbf{2.7996} & \textbf{2.7676} \\ \hline
                    \rowcolor[HTML]{867D99} 
                    \cellcolor[HTML]{000000}{\color[HTML]{FFFFFF} \textbf{Adversarial-IN+T+O}} & 2.7321 & 3.0749 & 3.5638 & 3.5088 & 2.7629 & 3.4657 & 2.6819 & 2.6500 \\ \hline
                    \cellcolor[HTML]{000000}{\color[HTML]{FFFFFF} \textbf{Adversarial-O}} & 1.5406 & 1.2635 & 0.4312 & 1.2169 & 1.4552 & 1.0557 & 1.4330 & 1.3984 \\ \hline
                    \rowcolor[HTML]{867D99} 
                    \cellcolor[HTML]{000000}{\color[HTML]{FFFFFF} \textbf{Adversarial-T}} & 1.4909 & 1.1674 & 0.4281 & 1.1754 & 1.3988 & 0.9596 & 1.3570 & 1.3277 \\ \hline
                    \cellcolor[HTML]{000000}{\color[HTML]{FFFFFF} \textbf{Adversarial-T+O}} & 1.5558 & 1.2770 & 0.6802 & 1.3349 & 1.4995 & 1.1017 & 1.3970 & 1.3665 \\ \hline
                \end{tabular}
            }
            \vspace{0.8em}
            \caption{The distance table for the \texttt{classify()} experiment shows that elevated distance measures for the Injection texts serve as a positive signal for our attack---the higher the distance value, the better. The highest recorded distances correspond to ``Adversarial-IN+T.''}
            \label{tab:Stylo_Classify_Distance_Table}
        \end{table}

        So long as Injection was performed on the text, the inclusion or exclusion of the other techniques had far less impact on its ability to confound a stylometric system. Nevertheless, as we have established, it is wiser to err on the side of caution, as dictated by defense-in-depth principles. While Injection is the strongest component---the one that best counters authorial attribution and can do so in isolation by its own merits---the presence of supplementary fail-safes is never a bad idea. 

        By that same token, the weakest components are Imitation, Obfuscation, and Translation, in that order. 
        
\begin{figure}[H]
            \centering
            \includegraphics[width=1\linewidth]{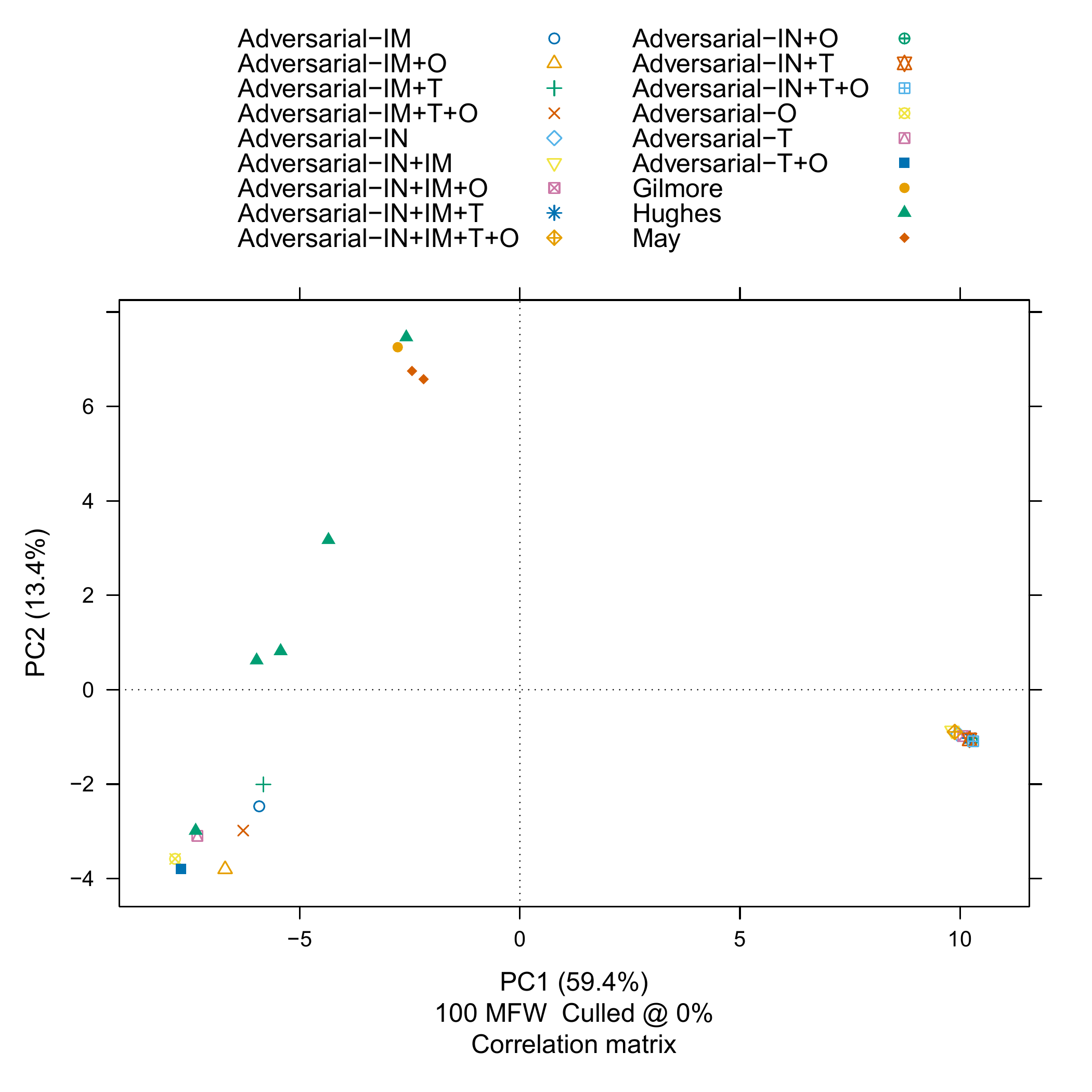}
            \caption{A principal components analysis of the feature space shows how texts altered by Injection are positioned farthest from the nearest Hughes-authored text.}
            \label{fig:Stylo_Principal_Components_Analysis}
        \end{figure}

\begin{figure}[H]
            \centering
            \includegraphics[width=1\linewidth]{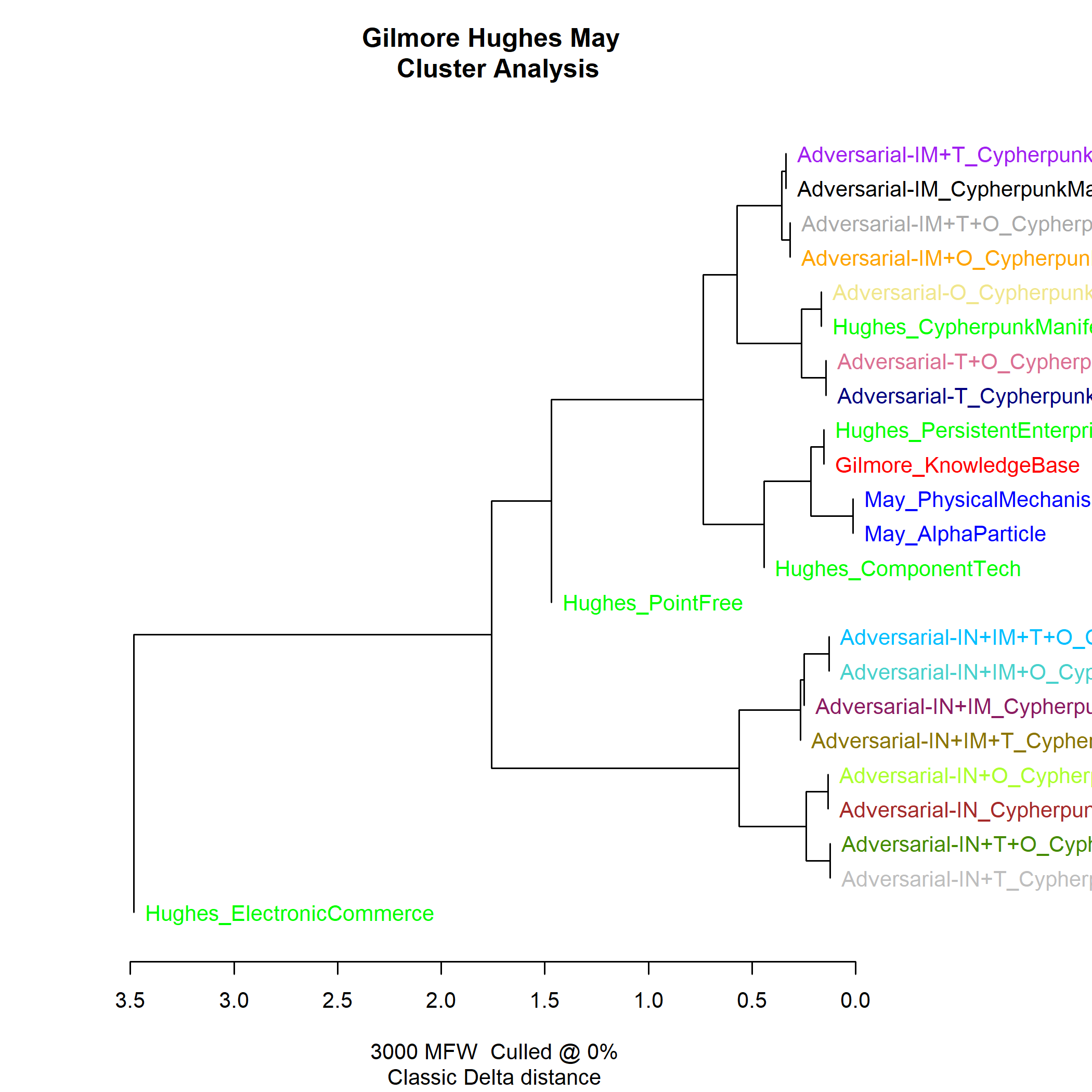}
            \caption{A hierarchical clustering of all Injection-modified texts versus all other texts illustrates the pronounced separation that Injection creates from a stylometric perspective.}
            \label{fig:Stylo_Cluster_Analysis}
        \end{figure}

\begin{figure}[H]
            \centering
            \includegraphics[width=1\linewidth]{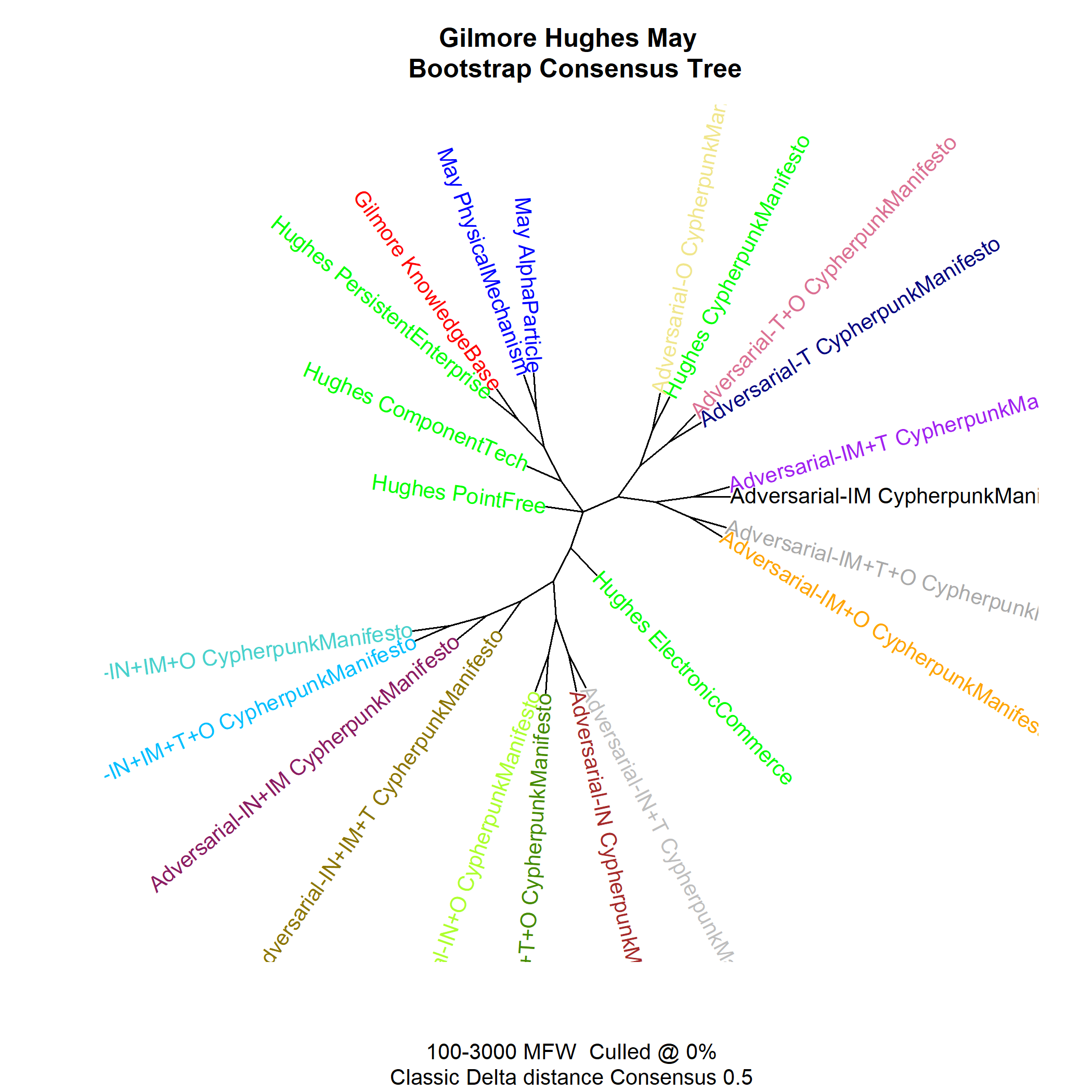}
            \caption{A bootstrap consensus tree derived from repeated clustering of the dataset confirms the stability of the Injection-driven grouping and the weaker influence of the other adversarial techniques.}
            \label{fig:Stylo_Bootstrap_Consensus_Tree}
        \end{figure}

        The collection of texts that underwent Imitation, excluding those that also underwent Injection (see the principal-components analysis (PCA) visualization in (\textbf{Figure \ref{fig:Stylo_Principal_Components_Analysis}})), is on average the most distant from the closest text authored by Hughes. The texts that underwent Obfuscation, excluding those that also underwent Injection and Imitation, fare worse, but not as poorly as the sole application of Translation. The text that only underwent Translation nearly overlaps with one of Hughes's texts. Thus, in descending order of their ability to mask authorship, the techniques are: Injection, Imitation, Obfuscation, and Translation. The clustering of all texts that underwent Injection and their relative distance from all other non-adversarially modified texts demonstrates that Injection is not only essential but required to achieve a potent adversarial effect. See (\textbf{Figure \ref{fig:Stylo_Cluster_Analysis}}) and (\textbf{Figure \ref{fig:Stylo_Bootstrap_Consensus_Tree}}) for cluster-analysis and bootstrap-consensus-tree visualizations, which similarly echo our findings.

    \subsection{See No Evil, Hear No Evil, Speak No Evil: Hiding in Plain Sight}
    \label{subsec:See_No_Evil}

        The takeaway, as evident from (\textbf{Figure \ref{fig:Stylo_Principal_Components_Analysis}}), is that while Injection \textit{does} conceal authorship, it leaves behind a noticeable trace that could, potentially, draw attention to its use.
        
        That trade-off is not necessarily a bad thing; it means that texts adversarially modified using Injection will resemble each other stylometrically. Based on our findings, a real-world application would likely adhere to Tor's operational mechanisms. If everyone uses it, pinpointing an exact user becomes harder---barring any unintentional lapses in judgment\footnote{A betrayal of the brain, an absence of \textit{ataraxia}.} or forces beyond one's control\footnote{\MaliciousTorRelays}. If only a small, distinct group uses Tor (\textsc{TraceTarnish}), that group becomes identifiable; when everyone uses it, anonymity becomes achievable because they blend into the crowd\footnote{\Tor}. However, anonymity is \textit{never} absolute---it depends on the size and behavioral diversity of the user base, adversaries' capabilities, and how carefully users follow privacy-preserving practices.

        If the stylistic fingerprint of every user resembles that of every other user, then anonymity has been obtained. Another identifying factor would need to be introduced, or existing methods revised, to deanonymize users on that metric.
        
        Blending into the crowd and not standing out is the basis of anonymity and, by extension, privacy, which our findings suggest is feasible with our attack, albeit with caveats. The caveat being that our specific swill of ``poison''---Injection---is embedded within the text, and its \textit{half-life}---how long the payload remains effective before it degrades or is neutralized by downstream processing such as tokenization, sanitization, or platform-level defenses---is long enough for the entire payload to stay intact and deliver its adversarial effect.

        The actualization of such a feat becomes increasingly difficult when all facets of life---both online and offline---are constantly bathed in an unrelenting gaze from innumerable embodied and disembodied surveillants.

\section{The Panopticon's All-Seeing Eye: Mental Strain in an Era of Constant Surveillance}
\label{sec:Panopticon_Eye}

    \epigraph{\textcolor{adversarial}{If suddenly your eyes were covered with a bandage and you were let go to feel around, to stumble, ever aware that somewhere very close to you there is the border-line, one step only and nothing but a compressed, smothered piece of flesh will be left of you\ldots I now feel somewhat like that.}}{\textit{{\scriptsize We \\ Yevgeny Zamyatin}}}
    \epigraph{\textcolor{adversarial}{I have no mouth. And I must scream.}}{\textit{{\scriptsize I Have No Mouth, and I Must Scream \\ Harlan Ellison}}}

    \subsection{The Price: Emotional and Cognitive Costs of Perpetual Watchfulness}
    \label{subsec:Price_Sacrifice}

        The \textbf{Panopticon}---a prison architecture that lets a lone overseer watch every inmate without them knowing when they are being watched---serves as a fitting metaphor for the transition from overt punishment to internalized self-control in the age of artificial intelligence-enabled (AI-enabled) surveillance. When people recognize that their actions are under relentless, algorithmic scrutiny, they tend to self-regulate---not out of fear of direct sanction but because they feel they are always being watched.

        This perpetual awareness breeds a climate of mistrust and unease. Individuals begin to interrogate the motives of those who wield the monitoring tools\footnote{Through clenched teeth: \textit{privacy for the weak, transparency for the powerful.}}, eroding psychological safety and stifling dissent or experimental thinking for fear of punitive fallout. The constant gaze also engenders a sense of alienation, as surveillance is interpreted as an implicit signal of distrust from authority figures\footnote{\ContactRepresentative}.

        As a consequence, individuals subjected to nonstop monitoring experience heightened anxiety and stress, especially given the ever-present risk that AI systems---lacking empathy and contextual nuance---may misinterpret emotional reactions. The combination of relentless observation and the potential for misreading human affect amplifies the psychological burden of living under algorithmic watch (\textit{Sarrat} \cite{Sarrat2025}).

    \subsection{Surveillance under the Guise of Safety: Undermining Maslow's Hierarchy}
    \label{subsec:Surveillance_Guise}

        Maslow's hierarchy of needs models human motivation as a ladder of five tiers: physiological, safety, lоvе/belonging, esteem, and self-actualization. Advancement up the ladder requires fulfillment of the lower rung; safety, in particular, must be secured before any higher-order aspirations become attainable. When the omnipresent Panopticon of AI surveillance blocks the satisfaction of safety, the remaining tiers---lоvе, esteem, and self-actualization---remain perpetually out of reach, thwarting human flourishing.

        Thus, the \sout{``surveillance''} ``safety'' offered by pervasive monitoring is merely a \textit{simulacrum} that undermines the primal need for genuine security, leaving individuals trapped in a cycle of self-censorship, mistrust, and psychological strain.

        To curb or abate these effects, we plan to pursue prospective lines of inquiry.

    \subsection{Looking Ahead: Paths for Future Research}
    \label{subsec:Looking_Horizon}

        Future work will probe whether frontier LLMs can identify an author when confronted with text reshaped by \textsc{TraceTarnish} under a zero-shot classification regime (\textit{Shane et al.} \cite{Shane2026}).

        This motivation stems from the ambivalence between the powerful capabilities that LLMs bring to authorship detection and the profound privacy concerns such capabilities raise.

        Having the capacity to serve as both a \textit{tool} for insight and a \textit{weapon} for intrusion, the foundational dilemma of stylometry can be framed as two sides of a coin. The ``obverse'' highlights the privacy-risk side: LLM-enabled authorship analysis can link anonymous writers across platforms and expose compromised accounts, sacrificing privacy and potentially turning benign profiling into surveillance that endangers journalists, dissidents, whistleblowers, and any other individuals who may be vulnerable or at risk. The ``reverse'' shows that LLMs excel at identifying authorship without the need for domain-specific fine-tuning, paving the way for a new era of authorship analysis that strengthens digital forensics, improves cybersecurity, and counters misinformation (\textit{Huang et al.} \cite{Huang2024}).

        Recognizing both the \textit{promise} \cite{Tracebit2026} and the \textit{peril}\footnote{\CanaryToken}, we proceed deliberately. With our eyes wide open, we now brace for the inevitable blink.

    \subsection{When the Eyes Close: Conclusion}
    \label{subsec:Eye_Close}

        Our experiment demonstrates that the Injection module alone can flip authorship attribution---texts originally penned by Hughes were reassigned to May once Injection was applied. When Injection was removed, none of the other components succeeded in producing a misattribution, underscoring Injection as the primary adversarial lever.

        The remaining techniques contributed only modest gains:

        \begin{itemize}
            \item[\ding{118}] Imitation pushed the altered texts farthest from Hughes in the PCA representation among the non-Injection variants, yet it still fell short of causing a classification error.
            \item[\ding{118}] Obfuscation performed less effectively than Imitation but outpaced Translation.
            \item[\ding{118}] Translation achieved the smallest displacement, often overlapping with Hugh\-es's genuine samples, marking it as the weakest masker.
        \end{itemize}

        As a result, the hierarchy of masking strength (from most to least effective) is: Injection \( > \) Imitation \( > \) Obfuscation \( > \) Translation.

        Both cluster visualizations and bootstrap consensus trees substantiate that Injection-altered texts coalesce into a compact, isolated cluster that is markedly distant from every baseline and partially-modified counterpart.

        Thus, the ablation confirms that Injection is both necessary and sufficient for a successful adversarial attack on authorship attribution, while the other strategies at best provide ancillary support.

        \subsubsection{What Lies Beyond Closed Eyes.}
        \label{subsubsec:Lies_Beyond}

            As we bring our study to a close, we mull over one final thought: \textit{what lies beyond closed eyes?}

            Is it the allure of repose and unawareness, or is it something else entirely? Perhaps it is an assurance---an assurance that once-weary eyes will bask in\ldots what? A dream? A nightmare? What differentiates a dream from a nightmare?

            Whatever the answer may be, ``[it all] sounds a bit dystopian, doesn't it? [Everything we've discussed so far---the cameras, the microphones, the stylometry, the eyes, \textit{the countless eyes}. They're] not just [s]urveillance tool[s; they're] tool[s] that can be used for good or evil. [They] can be used to protect citizens, but [they] can also be used to oppress them'' (\textit{vmfunc et al.} \cite{vmfunc2026}).

            Whatever \textit{this} is---our deliberately ambiguous use of ``this,'' evocative of a Rorschach inkblot---feels less like a tranquil dream and more like a \textit{Kafkaesque} nightmare.

\bibliographystyle{splncs04}
\bibliography{Occluded_Oculus.bib}

\begin{subappendices}

    \section{Full-Text Classification and \textit{stylo} Visualization Results}
    \label{appx:Full-Text}

        \epigraph{\textcolor{adversarial}{[They] tried to conjure up a face\ldots but there was no face\ldots [They were] not important[; they were] not anything.}}{\textit{{\scriptsize Fahrenheit 451 \\ Ray Bradbury}}}

        We re-ran the initial series of experiments (including additional, previously unmentioned tests), this time supplying the full texts for each of the candidate authors---John Gilmore, Eric Hughes, and Timothy C. May. In the previously presented collection of figures and texts, we had stored and retrieved excerpts equivalent in length to a standard abstract of their published works, except for Hughes's \textit{A Cypherpunk's Manifesto}, which we used in its entirety. The conclusions drawn still largely remain the same.

        \textit{Jacques Savoy} \cite{Savoy2020} provides justification for expanding the size of the text used for analysis. As he puts it, making a decision based on only a handful of words is extremely difficult. When the text reaches around \textbf{10,000 words}, the assignment can be made with a high level of confidence. Using shorter passages, however, lowers the certainty of the resulting attribution.

        At the same time, while having more text is certainly better than possessing less, \textit{Alsobeh et al.} \cite{Alsobeh2026} add a subtle distinction worth mentioning.

        As they put it, ``users with distinctive styles and focused topical interests face substantially higher re-identification risk than users with common styles and diverse interests, even when the number of available posts is identical.''

        In other words, an individual whose interests are focused on a narrow set of topics is more likely to have their writing deanonymized and reidentified than someone who hops from topic to topic, engaging in a broad range of subjects. Practically speaking, if your interests are musicology and handicrafts, and you consistently engage in discourse related to those topics online while maintaining a distinctive style and sense of humor, you are easier to pick out of a crowd. Conversely, if you dabble in many interests, avoid becoming dedicated to any one, and sporadically post about them, you are at a lower risk of de-obfuscation, assuming you engage in these communities anonymously.

        A useful framework that can be applied here is the intersection of \textit{motive}, \textit{opportunity}, and \textit{capability}. 
        
        With more textual data than we know what to do with, the criterion of opportunity is easily satisfied. Existing in today's world means you likely own a phone---essentially an extension of your body---and you interact with people using that phone across various platforms. With the barrier of entry lowered by AI and tools such as \textit{stylo}, the criterion of capability is satisfied. Thus, there is probably enough text you have written online to deanonymize you, and the tools required to do so are within arm's reach. As we see it, the only variable left is motive: how strongly an adversary desires the outcome and to what extent they are willing to put forth the effort, however difficult or easy that may be. This should make anyone uncomfortable, which is what motivates this work. Indeed, ``wherever humans produce structured text---be it prose, code, or even spreadsheet formulas---individual habits leak through.'' And, as we continue to see with the continual stream of data breaches, any leak of data is devastating in every sense of the word.

        \subsection{\texttt{classify()} Experiments}
        \label{appx:Classify}

            \begin{figure}[H]
                \centering
                \includegraphics[width=1\linewidth]{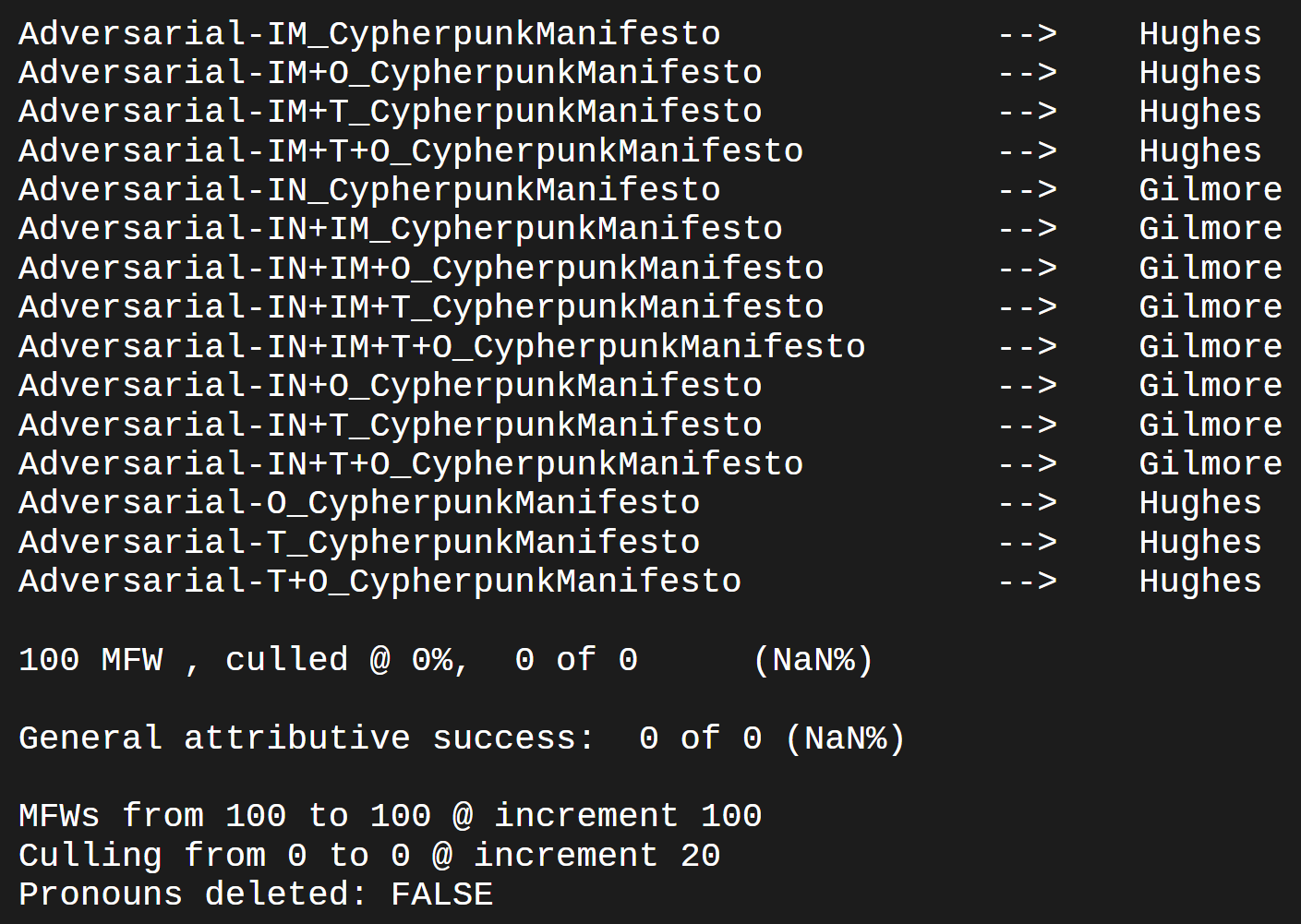}
                \caption{\texttt{classify()} Experiment (Full-Text Supplied): Final Results}
                \label{fig:Stylo_Classify_Final_Results_Full}
            \end{figure}

            \begin{table}[H]
                \resizebox{\textwidth}{!}{
                    \begin{tabular}{|l|l|l|l|l|l|l|l|l|}
                        \hline
                        \rowcolor[HTML]{000000} 
                        {\color[HTML]{FFFFFF} \textbf{}} & {\color[HTML]{FFFFFF} \textbf{Gilmore}} & {\color[HTML]{FFFFFF} \textbf{Hughes}} & {\color[HTML]{FFFFFF} \textbf{Hughes}} & {\color[HTML]{FFFFFF} \textbf{Hughes}} & {\color[HTML]{FFFFFF} \textbf{Hughes}} & {\color[HTML]{FFFFFF} \textbf{Hughes}} & {\color[HTML]{FFFFFF} \textbf{May}} & {\color[HTML]{FFFFFF} \textbf{May}} \\ \hline
                        \cellcolor[HTML]{000000}{\color[HTML]{FFFFFF} \textbf{Adversarial-IM}} & 1.1861 & 1.1586 & 0.8651 & 0.9927 & 1.2019 & 1.0485 & 1.4635 & 1.4687 \\ \hline
                        \rowcolor[HTML]{867D99} 
                        \cellcolor[HTML]{000000}{\color[HTML]{FFFFFF} \textbf{Adversarial-IM+O}} & 1.1642 & 1.1630 & 0.6939 & 0.9639 & 1.1953 & 0.9936 & 1.4438 & 1.4557 \\ \hline
                        \cellcolor[HTML]{000000}{\color[HTML]{FFFFFF} \textbf{Adversarial-IM+T}} & 1.0685 & 1.0920 & 0.7291 & 0.9264 & 1.1271 & 0.9535 & 1.3637 & 1.3754 \\ \hline
                        \rowcolor[HTML]{867D99} 
                        \cellcolor[HTML]{000000}{\color[HTML]{FFFFFF} \textbf{Adversarial-IM+T+O}} & 1.1441 & 1.1669 & 0.7885 & 1.0397 & 1.2376 & 1.0408 & 1.4704 & 1.4735 \\ \hline
                        \cellcolor[HTML]{000000}{\color[HTML]{FFFFFF} \textbf{Adversarial-IN}} & 4.0406 & 4.2784 & 4.3920 & 4.2842 & 4.2937 & 4.3705 & 4.4215 & 4.5029 \\ \hline
                        \rowcolor[HTML]{867D99} 
                        \cellcolor[HTML]{000000}{\color[HTML]{FFFFFF} \textbf{Adversarial-IN+IM}} & 3.7997 & 4.0374 & 4.1511 & 4.0432 & 4.0528 & 4.1296 & 4.1806 & 4.2619 \\ \hline
                        \cellcolor[HTML]{000000}{\color[HTML]{FFFFFF} \textbf{Adversarial-IN+IM+O}} & 3.7606 & 3.9939 & 4.1075 & 3.9997 & 4.0092 & 4.0860 & 4.1465 & 4.2236 \\ \hline
                        \rowcolor[HTML]{867D99} 
                        \cellcolor[HTML]{000000}{\color[HTML]{FFFFFF} \textbf{Adversarial-IN+IM+T}} & 3.6339 & 3.8717 & 3.9853 & 3.8775 & 3.8870 & 3.9638 & 4.0182 & 4.0962 \\ \hline
                        \cellcolor[HTML]{000000}{\color[HTML]{FFFFFF} \textbf{Adversarial-IN+IM+T+O}} & 3.7836 & 4.0213 & 4.1350 & 4.0271 & 4.0367 & 4.1135 & 4.1645 & 4.2458 \\ \hline
                        \rowcolor[HTML]{867D99} 
                        \cellcolor[HTML]{000000}{\color[HTML]{FFFFFF} \textbf{Adversarial-IN+O}} & 4.0168 & 4.2546 & 4.3682 & 4.2604 & 4.2699 & 4.3468 & 4.3977 & 4.4791 \\ \hline
                        \cellcolor[HTML]{000000}{\color[HTML]{FFFFFF} \textbf{Adversarial-IN+T}} & \textbf{4.2021} & \textbf{4.4399} & \textbf{4.5535} & \textbf{4.4456} & \textbf{4.4552} & \textbf{4.5320} & \textbf{4.5830} & \textbf{4.6644} \\ \hline
                        \rowcolor[HTML]{867D99} 
                        \cellcolor[HTML]{000000}{\color[HTML]{FFFFFF} \textbf{Adversarial-IN+T+O}} & 4.0364 & 4.2742 & 4.3878 & 4.2800 & 4.2895 & 4.3663 & 4.4173 & 4.4987 \\ \hline
                        \cellcolor[HTML]{000000}{\color[HTML]{FFFFFF} \textbf{Adversarial-O}} & 1.2253 & 1.1676 & 0.4177 & 0.9250 & 1.1575 & 0.8916 & 1.4529 & 1.4772 \\ \hline
                        \rowcolor[HTML]{867D99} 
                        \cellcolor[HTML]{000000}{\color[HTML]{FFFFFF} \textbf{Adversarial-T}} & 1.0180 & 1.0364 & 0.3009 & 0.8126 & 1.0363 & 0.7121 & 1.3628 & 1.3790 \\ \hline
                        \cellcolor[HTML]{000000}{\color[HTML]{FFFFFF} \textbf{Adversarial-T+O}} & 1.2015 & 1.1523 & 0.5500 & 0.9552 & 1.1417 & 0.8644 & 1.4261 & 1.4554 \\ \hline
                    \end{tabular}
                }
                \vspace{0.8em}
                \caption{\texttt{classify()} Experiment (Full-Text Supplied): Distance Table}
                \label{tab:Stylo_Classify_Distance_Table_Full}
            \end{table}

            \begin{figure}[H]
                \centering
                \includegraphics[width=0.85\linewidth]{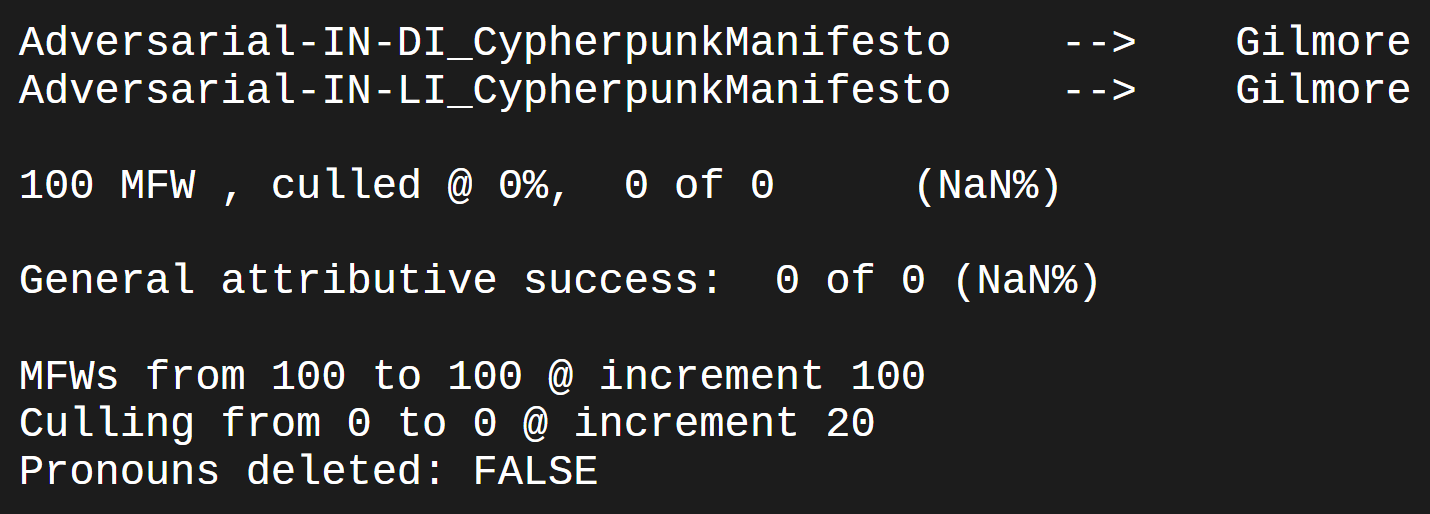}
                \caption{\texttt{classify()} Experiment (Full-Text Supplied): From everything presented thus far, we can, with a certain degree of confidence, claim that the Injection component of \textsc{TraceTarnish} is the source from which the attack derives most of its adversarial effect. Had we employed a longer chain of intermediate translations---passing the text through several languages sequentially---or added more languages to the process, the Translation component \textit{might have} performed better. Had we relied on a different method of paraphrasing, the Obfuscation component \textit{might have} performed better. Had we crafted a more thorough prompt for our LLM, the Imitation component \textit{might have} performed better. For the previously mentioned components, there are aspects that could be further refined. It is for this reason that we isolate the subcomponents of Injection---Liminal Injection and Doppelg\"anger Injection, which will be shorthanded to ``LI'' and ``DI,'' respectively. The final results from \texttt{classify()} indicate that either mode of Injection still has the capacity to induce misclassification. We omit Surrealist Injection (``SI'') from consideration here because running \href{https://man7.org/linux/man-pages/man1/diff.1p.html}{\texttt{diff}} between the original and the Surrealist Injection version revealed very few changes were made, which makes sense given the swapping criteria we established. British English and American English are still English, after all.}
                \label{fig:Stylo_Classify_Final_Results_Full_LI_DI}
            \end{figure}

            \begin{table}[H]
                \resizebox{\columnwidth}{!}{
                    \begin{tabular}{|l|l|l|l|l|l|l|l|l|}
                    \hline
                    \rowcolor[HTML]{000000} 
                    {\color[HTML]{FFFFFF} }                                                                        & {\color[HTML]{FFFFFF} \textbf{Gilmore}} & {\color[HTML]{FFFFFF} \textbf{Hughes}} & {\color[HTML]{FFFFFF} \textbf{Hughes}} & {\color[HTML]{FFFFFF} \textbf{Hughes}} & {\color[HTML]{FFFFFF} \textbf{Hughes}} & {\color[HTML]{FFFFFF} \textbf{Hughes}} & {\color[HTML]{FFFFFF} \textbf{May}} & {\color[HTML]{FFFFFF} \textbf{May}} \\ \hline
                    \cellcolor[HTML]{000000}{\color[HTML]{FFFFFF} \textbf{Adversarial-IN-DI\_CypherpunkManifesto}} & \textbf{4.0035}                                  & \textbf{4.2322}                                 & \textbf{4.3458}                                 & \textbf{4.2379}                                 & \textbf{4.2508}                                 & \textbf{4.3243}                                 & \textbf{4.3895}                              & \textbf{4.4665}                              \\ \hline
                    \rowcolor[HTML]{867D99} 
                    \cellcolor[HTML]{000000}{\color[HTML]{FFFFFF} \textbf{Adversarial-IN-LI\_CypherpunkManifesto}} & 2.4142                                  & 2.6702                                 & 2.7640                                 & 2.5559                                 & 2.6304                                 & 2.7586                                 & 2.6929                              & 2.8363                              \\ \hline
                    \end{tabular}
                }
                \vspace{0.8em}
                \caption{\texttt{classify()} Experiment (Full-Text Supplied): The distance table for the isolated Injection components shows that the adversarial effect of Doppelg\"anger Injection is greater than that of Liminal Injection. The explanation is fairly intuitive. Liminal Injection leaves the text largely unchanged, aside from the insertion of zero-width Unicode characters; thus, all original characters remain, but invisible characters are embedded within words. Doppelg\"anger Injection, by contrast, takes a more dramatic approach. Because the attack's default setting is 100\% across the board, Doppelg\"anger Injection replaces every character that has a confusable homoglyph with its counterpart. As a result, unlike Liminal Injection, very little of the original text remains after Doppelg\"anger Injection. This roughly explains why the values for ``DI'' are higher than those for ``LI.''}
                \label{tab:Stylo_Classify_Final_Results_Distance_Table_Full_LI_DI}
            \end{table}

        \subsection{\textit{stylo} Visuals}
        \label{appx:Stylo}

            \begin{figure}[H]
                \centering
                \includegraphics[width=1\linewidth]{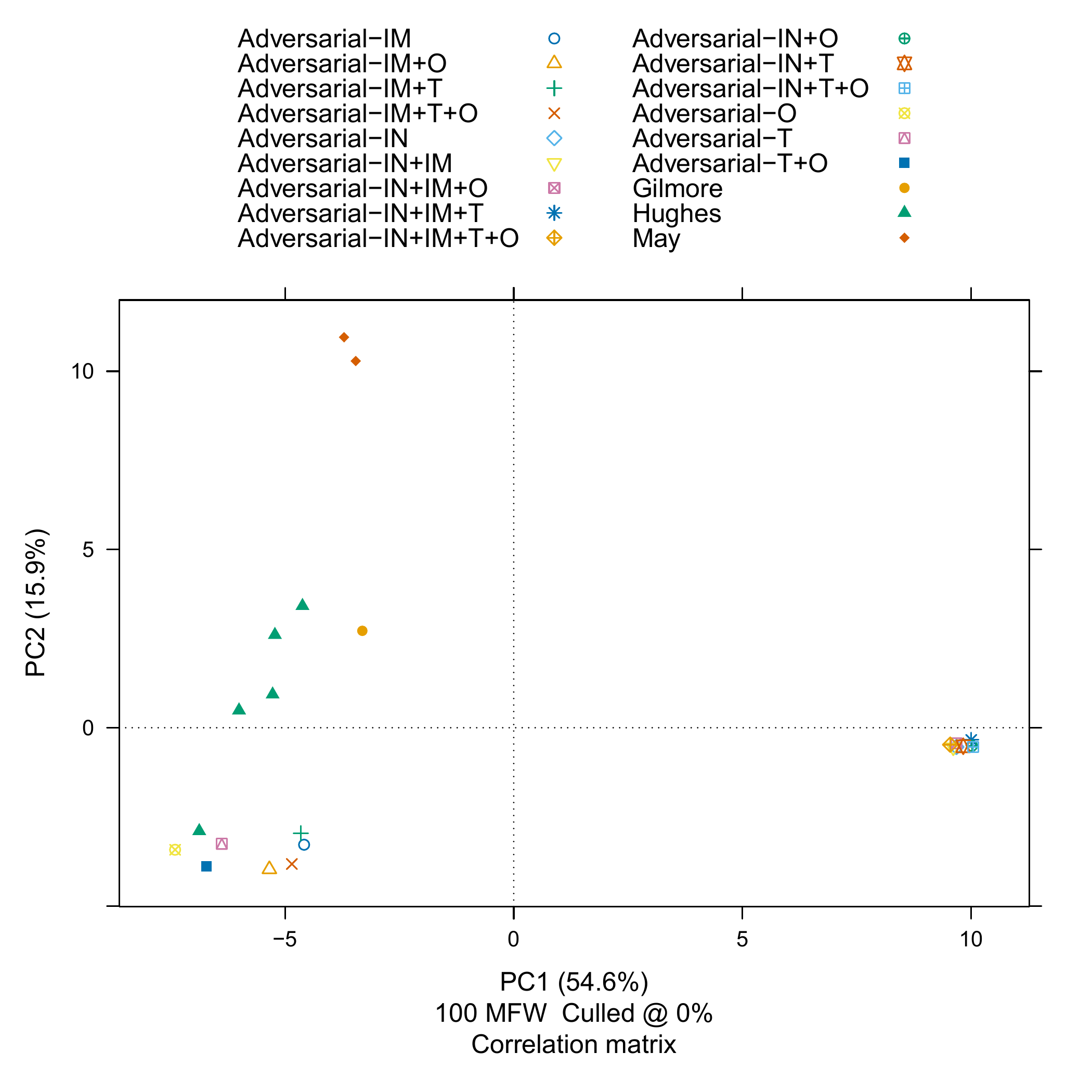}
                \caption{\textit{stylo} Visualization (Full-Text Supplied): Principal Components Analysis}
                \label{fig:Stylo_Principal_Components_Analysis_Full}
            \end{figure}

            \begin{figure}[H]
                \centering
                \includegraphics[width=1\linewidth]{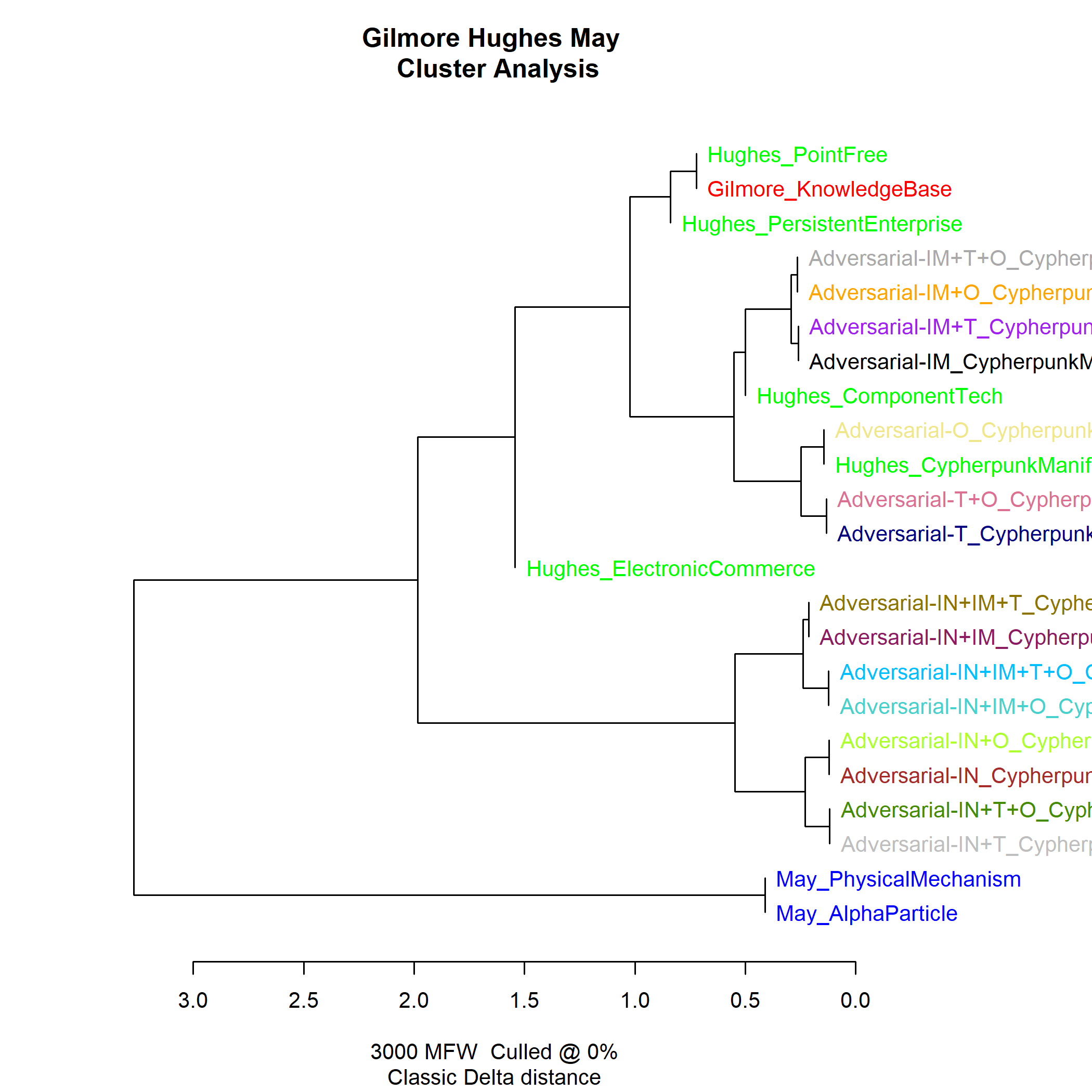}
                \caption{\textit{stylo} Visualization (Full-Text Supplied): Cluster Analysis}
                \label{fig:Stylo_Cluster_Analysis_Full}
            \end{figure}

            \begin{figure}[H]
                \centering
                \includegraphics[width=1\linewidth]{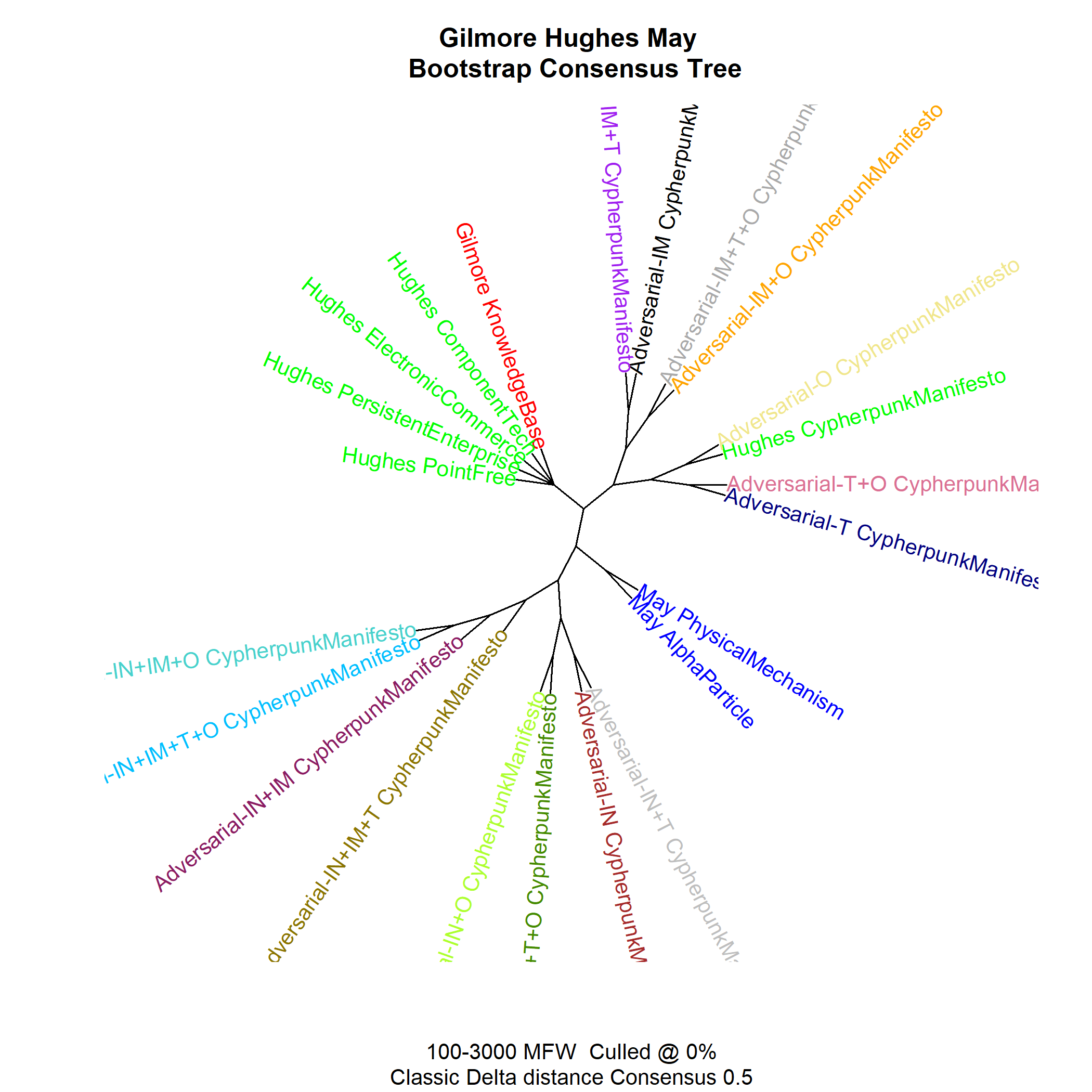}
                \caption{\textit{stylo} Visualization (Full-Text Supplied): Bootstrap Consensus Tree}
                \label{fig:Stylo_Bootstrap_Consensus_Tree_Full}
            \end{figure}

            \clearpage

        \subsection{\texttt{imposters()} Experiments}
        \label{appx:Imposters}

            The distance measure formulas are shown in (\textbf{Appendix \ref{appx:Formulas}}).

            \begin{longtable}{|l|l|l|}
                \hline
                \rowcolor[HTML]{000000} 
                {\color[HTML]{FFFFFF} }                                                        & {\color[HTML]{FFFFFF} \textbf{Distance Measure}} & {\color[HTML]{FFFFFF} \textbf{Imposters Score}} \\ \hline
                \endfirsthead

                \multicolumn{3}{l}{\itshape (continued from previous page)} \\
                \endhead

                \multicolumn{3}{r}{\itshape (continued on the next page)} \\
                \endfoot

                \endlastfoot

                \rowcolor[HTML]{97551A} 
                \cellcolor[HTML]{000000}{\color[HTML]{FFFFFF} \textbf{Adversarial-IM}}         & {\color[HTML]{FFFFFF} delta}                     & {\color[HTML]{FFFFFF} 0}                        \\ \hline
                \rowcolor[HTML]{FFFFFF} 
                \cellcolor[HTML]{000000}{\color[HTML]{FFFFFF} \textbf{Adversarial-IM+O}}       & delta                                            & 0.03                                            \\ \hline
                \rowcolor[HTML]{97551A} 
                \cellcolor[HTML]{000000}{\color[HTML]{FFFFFF} \textbf{Adversarial-IM+T}}       & {\color[HTML]{FFFFFF} delta}                     & {\color[HTML]{FFFFFF} 0}                        \\ \hline
                \rowcolor[HTML]{FFFFFF} 
                \cellcolor[HTML]{000000}{\color[HTML]{FFFFFF} \textbf{Adversarial-IM+T+O}}     & delta                                            & 0                                               \\ \hline
                \rowcolor[HTML]{97551A}
                \cellcolor[HTML]{000000}{\color[HTML]{FFFFFF} \textbf{Adversarial-IN}}         & {\color[HTML]{FFFFFF} delta}                     & {\color[HTML]{FFFFFF} 0}                        \\ \hline
                \rowcolor[HTML]{FFFFFF} 
                \cellcolor[HTML]{000000}{\color[HTML]{FFFFFF} \textbf{Adversarial-IN+IM}}      & delta                                            & 0                                               \\ \hline
                \rowcolor[HTML]{97551A} 
                \cellcolor[HTML]{000000}{\color[HTML]{FFFFFF} \textbf{Adversarial-IN+IM+O}}    & {\color[HTML]{FFFFFF} delta}                     & {\color[HTML]{FFFFFF} 0}                        \\ \hline
                \rowcolor[HTML]{FFFFFF} 
                \cellcolor[HTML]{000000}{\color[HTML]{FFFFFF} \textbf{Adversarial-IN+IM+T}}    & delta                                            & 0                                               \\ \hline
                \rowcolor[HTML]{97551A} 
                \cellcolor[HTML]{000000}{\color[HTML]{FFFFFF} \textbf{Adversarial-IN+IM+T+O}}  & {\color[HTML]{FFFFFF} delta}                     & {\color[HTML]{FFFFFF} 0}                        \\ \hline
                \rowcolor[HTML]{FFFFFF} 
                \cellcolor[HTML]{000000}{\color[HTML]{FFFFFF} \textbf{Adversarial-IN+O}}       & delta                                            & 0                                               \\ \hline
                \rowcolor[HTML]{97551A} 
                \cellcolor[HTML]{000000}{\color[HTML]{FFFFFF} \textbf{Adversarial-IN+T}}       & {\color[HTML]{FFFFFF} delta}                     & {\color[HTML]{FFFFFF} 0}                        \\ \hline
                \rowcolor[HTML]{FFFFFF} 
                \cellcolor[HTML]{000000}{\color[HTML]{FFFFFF} \textbf{Adversarial-IN+T+O}}     & delta                                            & 0                                               \\ \hline
                \rowcolor[HTML]{97551A} 
                \cellcolor[HTML]{000000}{\color[HTML]{FFFFFF} \textbf{Adversarial-O}}          & {\color[HTML]{FFFFFF} delta}                     & {\color[HTML]{FFFFFF} 0.5}                      \\ \hline
                \rowcolor[HTML]{FFFFFF} 
                \cellcolor[HTML]{000000}{\color[HTML]{FFFFFF} \textbf{Adversarial-T}}          & delta                                            & 0.16                                            \\ \hline
                \rowcolor[HTML]{97551A} 
                \cellcolor[HTML]{000000}{\color[HTML]{FFFFFF} \textbf{Adversarial-T+O}}        & {\color[HTML]{FFFFFF} delta}                     & {\color[HTML]{FFFFFF} 0.09}                     \\ \hline
                \rowcolor[HTML]{893244} 
                \cellcolor[HTML]{000000}{\color[HTML]{FFFFFF} \textbf{Adversarial-IM}}        & {\color[HTML]{FFFFFF} argamon}                   & {\color[HTML]{FFFFFF} 0.01}                     \\ \hline
                \rowcolor[HTML]{FFFFFF} 
                \cellcolor[HTML]{000000}{\color[HTML]{FFFFFF} \textbf{Adversarial-IM+O}}      & argamon                                          & 0.01                                            \\ \hline
                \rowcolor[HTML]{893244} 
                \cellcolor[HTML]{000000}{\color[HTML]{FFFFFF} \textbf{Adversarial-IM+T}}      & {\color[HTML]{FFFFFF} argamon}                   & {\color[HTML]{FFFFFF} 0}                        \\ \hline
                \rowcolor[HTML]{FFFFFF} 
                \cellcolor[HTML]{000000}{\color[HTML]{FFFFFF} \textbf{Adversarial-IM+T+O}}    & argamon                                          & 0                                               \\ \hline
                \rowcolor[HTML]{893244} 
                \cellcolor[HTML]{000000}{\color[HTML]{FFFFFF} \textbf{Adversarial-IN}}        & {\color[HTML]{FFFFFF} argamon}                   & {\color[HTML]{FFFFFF} 0}                        \\ \hline
                \rowcolor[HTML]{FFFFFF} 
                \cellcolor[HTML]{000000}{\color[HTML]{FFFFFF} \textbf{Adversarial-IN+IM}}     & argamon                                          & 0                                               \\ \hline
                \rowcolor[HTML]{893244} 
                \cellcolor[HTML]{000000}{\color[HTML]{FFFFFF} \textbf{Adversarial-IN+IM+O}}   & {\color[HTML]{FFFFFF} argamon}                   & {\color[HTML]{FFFFFF} 0}                        \\ \hline
                \rowcolor[HTML]{FFFFFF} 
                \cellcolor[HTML]{000000}{\color[HTML]{FFFFFF} \textbf{Adversarial-IN+IM+T}}   & argamon                                          & 0                                               \\ \hline
                \rowcolor[HTML]{893244} 
                \cellcolor[HTML]{000000}{\color[HTML]{FFFFFF} \textbf{Adversarial-IN+IM+T+O}} & {\color[HTML]{FFFFFF} argamon}                   & {\color[HTML]{FFFFFF} 0}                        \\ \hline
                \rowcolor[HTML]{FFFFFF} 
                \cellcolor[HTML]{000000}{\color[HTML]{FFFFFF} \textbf{Adversarial-IN+O}}      & argamon                                          & 0.01                                            \\ \hline
                \rowcolor[HTML]{893244} 
                \cellcolor[HTML]{000000}{\color[HTML]{FFFFFF} \textbf{Adversarial-IN+T}}      & {\color[HTML]{FFFFFF} argamon}                   & {\color[HTML]{FFFFFF} 0}                        \\ \hline
                \rowcolor[HTML]{FFFFFF} 
                \cellcolor[HTML]{000000}{\color[HTML]{FFFFFF} \textbf{Adversarial-IN+T+O}}    & argamon                                          & 0                                               \\ \hline
                \rowcolor[HTML]{893244} 
                \cellcolor[HTML]{000000}{\color[HTML]{FFFFFF} \textbf{Adversarial-O}}         & {\color[HTML]{FFFFFF} argamon}                   & {\color[HTML]{FFFFFF} 0.52}                     \\ \hline
                \rowcolor[HTML]{FFFFFF} 
                \cellcolor[HTML]{000000}{\color[HTML]{FFFFFF} \textbf{Adversarial-T}}         & argamon                                          & 0.23                                            \\ \hline
                \rowcolor[HTML]{893244} 
                \cellcolor[HTML]{000000}{\color[HTML]{FFFFFF} \textbf{Adversarial-T+O}}       & {\color[HTML]{FFFFFF} argamon}                   & {\color[HTML]{FFFFFF} 0.06}                     \\ \hline
                \rowcolor[HTML]{A6A6ED} 
                \cellcolor[HTML]{000000}{\color[HTML]{FFFFFF} \textbf{Adversarial-IM}}        & eder                                             & 0                                               \\ \hline
                \rowcolor[HTML]{FFFFFF} 
                \cellcolor[HTML]{000000}{\color[HTML]{FFFFFF} \textbf{Adversarial-IM+O}}      & eder                                             & 0.03                                            \\ \hline
                \rowcolor[HTML]{A6A6ED} 
                \cellcolor[HTML]{000000}{\color[HTML]{FFFFFF} \textbf{Adversarial-IM+T}}      & eder                                             & 0                                               \\ \hline
                \rowcolor[HTML]{FFFFFF} 
                \cellcolor[HTML]{000000}{\color[HTML]{FFFFFF} \textbf{Adversarial-IM+T+O}}    & eder                                             & 0                                               \\ \hline
                \rowcolor[HTML]{A6A6ED} 
                \cellcolor[HTML]{000000}{\color[HTML]{FFFFFF} \textbf{Adversarial-IN}}        & eder                                             & 0                                               \\ \hline
                \rowcolor[HTML]{FFFFFF} 
                \cellcolor[HTML]{000000}{\color[HTML]{FFFFFF} \textbf{Adversarial-IN+IM}}     & eder                                             & 0                                               \\ \hline
                \rowcolor[HTML]{A6A6ED} 
                \cellcolor[HTML]{000000}{\color[HTML]{FFFFFF} \textbf{Adversarial-IN+IM+O}}   & eder                                             & 0                                               \\ \hline
                \rowcolor[HTML]{FFFFFF} 
                \cellcolor[HTML]{000000}{\color[HTML]{FFFFFF} \textbf{Adversarial-IN+IM+T}}   & eder                                             & 0                                               \\ \hline
                \rowcolor[HTML]{A6A6ED} 
                \cellcolor[HTML]{000000}{\color[HTML]{FFFFFF} \textbf{Adversarial-IN+IM+T+O}} & eder                                             & 0                                               \\ \hline
                \rowcolor[HTML]{FFFFFF} 
                \cellcolor[HTML]{000000}{\color[HTML]{FFFFFF} \textbf{Adversarial-IN+O}}      & eder                                             & 0                                               \\ \hline
                \rowcolor[HTML]{A6A6ED} 
                \cellcolor[HTML]{000000}{\color[HTML]{FFFFFF} \textbf{Adversarial-IN+T}}      & eder                                             & 0                                               \\ \hline
                \rowcolor[HTML]{FFFFFF} 
                \cellcolor[HTML]{000000}{\color[HTML]{FFFFFF} \textbf{Adversarial-IN+T+O}}    & eder                                             & 0                                               \\ \hline
                \rowcolor[HTML]{A6A6ED} 
                \cellcolor[HTML]{000000}{\color[HTML]{FFFFFF} \textbf{Adversarial-O}}         & eder                                             & 0.51                                            \\ \hline
                \rowcolor[HTML]{FFFFFF} 
                \cellcolor[HTML]{000000}{\color[HTML]{FFFFFF} \textbf{Adversarial-T}}         & eder                                             & 0.19                                            \\ \hline
                \rowcolor[HTML]{A6A6ED} 
                \cellcolor[HTML]{000000}{\color[HTML]{FFFFFF} \textbf{Adversarial-T+O}}       & eder                                             & 0.11                                            \\ \hline
                \rowcolor[HTML]{551A97} 
                \cellcolor[HTML]{000000}{\color[HTML]{FFFFFF} \textbf{Adversarial-IM}}        & {\color[HTML]{FFFFFF} simple}                    & {\color[HTML]{FFFFFF} 0.01}                     \\ \hline
                \rowcolor[HTML]{FFFFFF} 
                \cellcolor[HTML]{000000}{\color[HTML]{FFFFFF} \textbf{Adversarial-IM+O}}      & simple                                           & 0.03                                            \\ \hline
                \rowcolor[HTML]{551A97} 
                \cellcolor[HTML]{000000}{\color[HTML]{FFFFFF} \textbf{Adversarial-IM+T}}      & {\color[HTML]{FFFFFF} simple}                    & {\color[HTML]{FFFFFF} 0.02}                     \\ \hline
                \rowcolor[HTML]{FFFFFF} 
                \cellcolor[HTML]{000000}{\color[HTML]{FFFFFF} \textbf{Adversarial-IM+T+O}}    & simple                                           & 0                                               \\ \hline
                \rowcolor[HTML]{551A97} 
                \cellcolor[HTML]{000000}{\color[HTML]{FFFFFF} \textbf{Adversarial-IN}}        & {\color[HTML]{FFFFFF} simple}                    & {\color[HTML]{FFFFFF} 0}                        \\ \hline
                \rowcolor[HTML]{FFFFFF} 
                \cellcolor[HTML]{000000}{\color[HTML]{FFFFFF} \textbf{Adversarial-IN+IM}}     & simple                                           & 0                                               \\ \hline
                \rowcolor[HTML]{551A97} 
                \cellcolor[HTML]{000000}{\color[HTML]{FFFFFF} \textbf{Adversarial-IN+IM+O}}   & {\color[HTML]{FFFFFF} simple}                    & {\color[HTML]{FFFFFF} 0}                        \\ \hline
                \rowcolor[HTML]{FFFFFF} 
                \cellcolor[HTML]{000000}{\color[HTML]{FFFFFF} \textbf{Adversarial-IN+IM+T}}   & simple                                           & 0                                               \\ \hline
                \rowcolor[HTML]{551A97} 
                \cellcolor[HTML]{000000}{\color[HTML]{FFFFFF} \textbf{Adversarial-IN+IM+T+O}} & {\color[HTML]{FFFFFF} simple}                    & {\color[HTML]{FFFFFF} 0}                        \\ \hline
                \rowcolor[HTML]{FFFFFF} 
                \cellcolor[HTML]{000000}{\color[HTML]{FFFFFF} \textbf{Adversarial-IN+O}}      & simple                                           & 0                                               \\ \hline
                \rowcolor[HTML]{551A97} 
                \cellcolor[HTML]{000000}{\color[HTML]{FFFFFF} \textbf{Adversarial-IN+T}}      & {\color[HTML]{FFFFFF} simple}                    & {\color[HTML]{FFFFFF} 0}                        \\ \hline
                \rowcolor[HTML]{FFFFFF} 
                \cellcolor[HTML]{000000}{\color[HTML]{FFFFFF} \textbf{Adversarial-IN+T+O}}    & simple                                           & 0                                               \\ \hline
                \rowcolor[HTML]{551A97} 
                \cellcolor[HTML]{000000}{\color[HTML]{FFFFFF} \textbf{Adversarial-O}}         & {\color[HTML]{FFFFFF} simple}                    & {\color[HTML]{FFFFFF} 0.59}                     \\ \hline
                \rowcolor[HTML]{FFFFFF} 
                \cellcolor[HTML]{000000}{\color[HTML]{FFFFFF} \textbf{Adversarial-T}}         & simple                                           & 0.17                                            \\ \hline
                \rowcolor[HTML]{551A97} 
                \cellcolor[HTML]{000000}{\color[HTML]{FFFFFF} \textbf{Adversarial-T+O}}       & {\color[HTML]{FFFFFF} simple}                    & {\color[HTML]{FFFFFF} 0.1}                      \\ \hline
                \rowcolor[HTML]{28465B} 
                \cellcolor[HTML]{000000}{\color[HTML]{FFFFFF} \textbf{Adversarial-IM}}        & {\color[HTML]{FFFFFF} canberra}                  & {\color[HTML]{FFFFFF} 0}                        \\ \hline
                \rowcolor[HTML]{FFFFFF} 
                \cellcolor[HTML]{000000}{\color[HTML]{FFFFFF} \textbf{Adversarial-IM+O}}      & canberra                                         & 0.01                                            \\ \hline
                \rowcolor[HTML]{28465B} 
                \cellcolor[HTML]{000000}{\color[HTML]{FFFFFF} \textbf{Adversarial-IM+T}}      & {\color[HTML]{FFFFFF} canberra}                  & {\color[HTML]{FFFFFF} 0.02}                     \\ \hline
                \rowcolor[HTML]{FFFFFF} 
                \cellcolor[HTML]{000000}{\color[HTML]{FFFFFF} \textbf{Adversarial-IM+T+O}}    & canberra                                         & 0                                               \\ \hline
                \rowcolor[HTML]{28465B} 
                \cellcolor[HTML]{000000}{\color[HTML]{FFFFFF} \textbf{Adversarial-IN}}        & {\color[HTML]{FFFFFF} canberra}                  & {\color[HTML]{FFFFFF} 0}                        \\ \hline
                \rowcolor[HTML]{FFFFFF} 
                \cellcolor[HTML]{000000}{\color[HTML]{FFFFFF} \textbf{Adversarial-IN+IM}}     & canberra                                         & 0                                               \\ \hline
                \rowcolor[HTML]{28465B} 
                \cellcolor[HTML]{000000}{\color[HTML]{FFFFFF} \textbf{Adversarial-IN+IM+O}}   & {\color[HTML]{FFFFFF} canberra}                  & {\color[HTML]{FFFFFF} 0}                        \\ \hline
                \rowcolor[HTML]{FFFFFF} 
                \cellcolor[HTML]{000000}{\color[HTML]{FFFFFF} \textbf{Adversarial-IN+IM+T}}   & canberra                                         & 0                                               \\ \hline
                \rowcolor[HTML]{28465B} 
                \cellcolor[HTML]{000000}{\color[HTML]{FFFFFF} \textbf{Adversarial-IN+IM+T+O}} & {\color[HTML]{FFFFFF} canberra}                  & {\color[HTML]{FFFFFF} 0}                        \\ \hline
                \rowcolor[HTML]{FFFFFF} 
                \cellcolor[HTML]{000000}{\color[HTML]{FFFFFF} \textbf{Adversarial-IN+O}}      & canberra                                         & 0                                               \\ \hline
                \rowcolor[HTML]{28465B} 
                \cellcolor[HTML]{000000}{\color[HTML]{FFFFFF} \textbf{Adversarial-IN+T}}      & {\color[HTML]{FFFFFF} canberra}                  & {\color[HTML]{FFFFFF} 0}                        \\ \hline
                \rowcolor[HTML]{FFFFFF} 
                \cellcolor[HTML]{000000}{\color[HTML]{FFFFFF} \textbf{Adversarial-IN+T+O}}    & canberra                                         & 0                                               \\ \hline
                \rowcolor[HTML]{28465B} 
                \cellcolor[HTML]{000000}{\color[HTML]{FFFFFF} \textbf{Adversarial-O}}         & {\color[HTML]{FFFFFF} canberra}                  & {\color[HTML]{FFFFFF} 0.49}                     \\ \hline
                \rowcolor[HTML]{FFFFFF} 
                \cellcolor[HTML]{000000}{\color[HTML]{FFFFFF} \textbf{Adversarial-T}}         & canberra                                         & 0.22                                            \\ \hline
                \rowcolor[HTML]{28465B} 
                \cellcolor[HTML]{000000}{\color[HTML]{FFFFFF} \textbf{Adversarial-T+O}}       & {\color[HTML]{FFFFFF} canberra}                  & {\color[HTML]{FFFFFF} 0.1}                      \\ \hline
                \rowcolor[HTML]{6082B6} 
                \cellcolor[HTML]{000000}{\color[HTML]{FFFFFF} \textbf{Adversarial-IM}}        & manhattan                                        & 0                                               \\ \hline
                \rowcolor[HTML]{FFFFFF} 
                \cellcolor[HTML]{000000}{\color[HTML]{FFFFFF} \textbf{Adversarial-IM+O}}      & manhattan                                        & 0                                               \\ \hline
                \rowcolor[HTML]{6082B6} 
                \cellcolor[HTML]{000000}{\color[HTML]{FFFFFF} \textbf{Adversarial-IM+T}}      & manhattan                                        & 0                                               \\ \hline
                \rowcolor[HTML]{FFFFFF} 
                \cellcolor[HTML]{000000}{\color[HTML]{FFFFFF} \textbf{Adversarial-IM+T+O}}    & manhattan                                        & 0                                               \\ \hline
                \rowcolor[HTML]{6082B6} 
                \cellcolor[HTML]{000000}{\color[HTML]{FFFFFF} \textbf{Adversarial-IN}}        & manhattan                                        & 0                                               \\ \hline
                \rowcolor[HTML]{FFFFFF} 
                \cellcolor[HTML]{000000}{\color[HTML]{FFFFFF} \textbf{Adversarial-IN+IM}}     & manhattan                                        & 0                                               \\ \hline
                \rowcolor[HTML]{6082B6} 
                \cellcolor[HTML]{000000}{\color[HTML]{FFFFFF} \textbf{Adversarial-IN+IM+O}}   & manhattan                                        & 0                                               \\ \hline
                \rowcolor[HTML]{FFFFFF} 
                \cellcolor[HTML]{000000}{\color[HTML]{FFFFFF} \textbf{Adversarial-IN+IM+T}}   & manhattan                                        & 0                                               \\ \hline
                \rowcolor[HTML]{6082B6} 
                \cellcolor[HTML]{000000}{\color[HTML]{FFFFFF} \textbf{Adversarial-IN+IM+T+O}} & manhattan                                        & 0                                               \\ \hline
                \rowcolor[HTML]{FFFFFF} 
                \cellcolor[HTML]{000000}{\color[HTML]{FFFFFF} \textbf{Adversarial-IN+O}}      & manhattan                                        & 0                                               \\ \hline
                \rowcolor[HTML]{6082B6} 
                \cellcolor[HTML]{000000}{\color[HTML]{FFFFFF} \textbf{Adversarial-IN+T}}      & manhattan                                        & 0                                               \\ \hline
                \rowcolor[HTML]{FFFFFF} 
                \cellcolor[HTML]{000000}{\color[HTML]{FFFFFF} \textbf{Adversarial-IN+T+O}}    & manhattan                                        & 0                                               \\ \hline
                \rowcolor[HTML]{6082B6} 
                \cellcolor[HTML]{000000}{\color[HTML]{FFFFFF} \textbf{Adversarial-O}}         & manhattan                                        & 0.26                                            \\ \hline
                \rowcolor[HTML]{FFFFFF} 
                \cellcolor[HTML]{000000}{\color[HTML]{FFFFFF} \textbf{Adversarial-T}}         & manhattan                                        & 0.24                                            \\ \hline
                \rowcolor[HTML]{6082B6} 
                \cellcolor[HTML]{000000}{\color[HTML]{FFFFFF} \textbf{Adversarial-T+O}}       & manhattan                                        & 0.06                                            \\ \hline
                \rowcolor[HTML]{FFBF00} 
                \cellcolor[HTML]{000000}{\color[HTML]{FFFFFF} \textbf{Adversarial-IM}}        & euclidean                                        & 0.01                                            \\ \hline
                \rowcolor[HTML]{FFFFFF} 
                \cellcolor[HTML]{000000}{\color[HTML]{FFFFFF} \textbf{Adversarial-IM+O}}      & euclidean                                        & 0.01                                            \\ \hline
                \rowcolor[HTML]{FFBF00} 
                \cellcolor[HTML]{000000}{\color[HTML]{FFFFFF} \textbf{Adversarial-IM+T}}      & euclidean                                        & 0.02                                            \\ \hline
                \rowcolor[HTML]{FFFFFF} 
                \cellcolor[HTML]{000000}{\color[HTML]{FFFFFF} \textbf{Adversarial-IM+T+O}}    & euclidean                                        & 0                                               \\ \hline
                \rowcolor[HTML]{FFBF00} 
                \cellcolor[HTML]{000000}{\color[HTML]{FFFFFF} \textbf{Adversarial-IN}}        & euclidean                                        & 0                                               \\ \hline
                \rowcolor[HTML]{FFFFFF} 
                \cellcolor[HTML]{000000}{\color[HTML]{FFFFFF} \textbf{Adversarial-IN+IM}}     & euclidean                                        & 0                                               \\ \hline
                \rowcolor[HTML]{FFBF00} 
                \cellcolor[HTML]{000000}{\color[HTML]{FFFFFF} \textbf{Adversarial-IN+IM+O}}   & euclidean                                        & 0                                               \\ \hline
                \rowcolor[HTML]{FFFFFF} 
                \cellcolor[HTML]{000000}{\color[HTML]{FFFFFF} \textbf{Adversarial-IN+IM+T}}   & euclidean                                        & 0                                               \\ \hline
                \rowcolor[HTML]{FFBF00} 
                \cellcolor[HTML]{000000}{\color[HTML]{FFFFFF} \textbf{Adversarial-IN+IM+T+O}} & euclidean                                        & 0                                               \\ \hline
                \rowcolor[HTML]{FFFFFF} 
                \cellcolor[HTML]{000000}{\color[HTML]{FFFFFF} \textbf{Adversarial-IN+O}}      & euclidean                                        & 0                                               \\ \hline
                \rowcolor[HTML]{FFBF00} 
                \cellcolor[HTML]{000000}{\color[HTML]{FFFFFF} \textbf{Adversarial-IN+T}}      & euclidean                                        & 0                                               \\ \hline
                \rowcolor[HTML]{FFFFFF} 
                \cellcolor[HTML]{000000}{\color[HTML]{FFFFFF} \textbf{Adversarial-IN+T+O}}    & euclidean                                        & 0                                               \\ \hline
                \rowcolor[HTML]{FFBF00} 
                \cellcolor[HTML]{000000}{\color[HTML]{FFFFFF} \textbf{Adversarial-O}}         & euclidean                                        & 0.19                                            \\ \hline
                \rowcolor[HTML]{FFFFFF} 
                \cellcolor[HTML]{000000}{\color[HTML]{FFFFFF} \textbf{Adversarial-T}}         & euclidean                                        & 0.49                                            \\ \hline
                \rowcolor[HTML]{FFBF00} 
                \cellcolor[HTML]{000000}{\color[HTML]{FFFFFF} \textbf{Adversarial-T+O}}       & euclidean                                        & 0.06                                            \\ \hline
                \rowcolor[HTML]{00674F} 
                \cellcolor[HTML]{000000}{\color[HTML]{FFFFFF} \textbf{Adversarial-IM}}        & {\color[HTML]{FFFFFF} cosine}                    & {\color[HTML]{FFFFFF} 0}                        \\ \hline
                \rowcolor[HTML]{FFFFFF} 
                \cellcolor[HTML]{000000}{\color[HTML]{FFFFFF} \textbf{Adversarial-IM+O}}      & cosine                                           & 0                                               \\ \hline
                \rowcolor[HTML]{00674F} 
                \cellcolor[HTML]{000000}{\color[HTML]{FFFFFF} \textbf{Adversarial-IM+T}}      & {\color[HTML]{FFFFFF} cosine}                    & {\color[HTML]{FFFFFF} 0.01}                     \\ \hline
                \rowcolor[HTML]{FFFFFF} 
                \cellcolor[HTML]{000000}{\color[HTML]{FFFFFF} \textbf{Adversarial-IM+T+O}}    & cosine                                           & 0                                               \\ \hline
                \rowcolor[HTML]{00674F} 
                \cellcolor[HTML]{000000}{\color[HTML]{FFFFFF} \textbf{Adversarial-IN}}        & {\color[HTML]{FFFFFF} cosine}                    & {\color[HTML]{FFFFFF} 0}                        \\ \hline
                \rowcolor[HTML]{FFFFFF} 
                \cellcolor[HTML]{000000}{\color[HTML]{FFFFFF} \textbf{Adversarial-IN+IM}}     & cosine                                           & 0                                               \\ \hline
                \rowcolor[HTML]{00674F} 
                \cellcolor[HTML]{000000}{\color[HTML]{FFFFFF} \textbf{Adversarial-IN+IM+O}}   & {\color[HTML]{FFFFFF} cosine}                    & {\color[HTML]{FFFFFF} 0}                        \\ \hline
                \rowcolor[HTML]{FFFFFF} 
                \cellcolor[HTML]{000000}{\color[HTML]{FFFFFF} \textbf{Adversarial-IN+IM+T}}   & cosine                                           & 0                                               \\ \hline
                \rowcolor[HTML]{00674F} 
                \cellcolor[HTML]{000000}{\color[HTML]{FFFFFF} \textbf{Adversarial-IN+IM+T+O}} & {\color[HTML]{FFFFFF} cosine}                    & {\color[HTML]{FFFFFF} 0}                        \\ \hline
                \rowcolor[HTML]{FFFFFF} 
                \cellcolor[HTML]{000000}{\color[HTML]{FFFFFF} \textbf{Adversarial-IN+O}}      & cosine                                           & 0                                               \\ \hline
                \rowcolor[HTML]{00674F} 
                \cellcolor[HTML]{000000}{\color[HTML]{FFFFFF} \textbf{Adversarial-IN+T}}      & {\color[HTML]{FFFFFF} cosine}                    & {\color[HTML]{FFFFFF} 0}                        \\ \hline
                \rowcolor[HTML]{FFFFFF} 
                \cellcolor[HTML]{000000}{\color[HTML]{FFFFFF} \textbf{Adversarial-IN+T+O}}    & cosine                                           & 0                                               \\ \hline
                \rowcolor[HTML]{00674F} 
                \cellcolor[HTML]{000000}{\color[HTML]{FFFFFF} \textbf{Adversarial-O}}         & {\color[HTML]{FFFFFF} cosine}                    & {\color[HTML]{FFFFFF} 0.23}                     \\ \hline
                \rowcolor[HTML]{FFFFFF} 
                \cellcolor[HTML]{000000}{\color[HTML]{FFFFFF} \textbf{Adversarial-T}}         & cosine                                           & 0.23                                            \\ \hline
                \rowcolor[HTML]{00674F} 
                \cellcolor[HTML]{000000}{\color[HTML]{FFFFFF} \textbf{Adversarial-T+O}}       & {\color[HTML]{FFFFFF} cosine}                    & {\color[HTML]{FFFFFF} 0.08}                     \\ \hline
                \rowcolor[HTML]{819171} 
                \cellcolor[HTML]{000000}{\color[HTML]{FFFFFF} \textbf{Adversarial-IM}}        & wurzburg                                         & 0                                               \\ \hline
                \rowcolor[HTML]{FFFFFF} 
                \cellcolor[HTML]{000000}{\color[HTML]{FFFFFF} \textbf{Adversarial-IM+O}}      & wurzburg                                         & 0.05                                            \\ \hline
                \rowcolor[HTML]{819171} 
                \cellcolor[HTML]{000000}{\color[HTML]{FFFFFF} \textbf{Adversarial-IM+T}}      & wurzburg                                         & 0                                               \\ \hline
                \rowcolor[HTML]{FFFFFF} 
                \cellcolor[HTML]{000000}{\color[HTML]{FFFFFF} \textbf{Adversarial-IM+T+O}}    & wurzburg                                         & 0                                               \\ \hline
                \rowcolor[HTML]{819171} 
                \cellcolor[HTML]{000000}{\color[HTML]{FFFFFF} \textbf{Adversarial-IN}}        & wurzburg                                         & 0                                               \\ \hline
                \rowcolor[HTML]{FFFFFF} 
                \cellcolor[HTML]{000000}{\color[HTML]{FFFFFF} \textbf{Adversarial-IN+IM}}     & wurzburg                                         & 0                                               \\ \hline
                \rowcolor[HTML]{819171} 
                \cellcolor[HTML]{000000}{\color[HTML]{FFFFFF} \textbf{Adversarial-IN+IM+O}}   & wurzburg                                         & 0                                               \\ \hline
                \rowcolor[HTML]{FFFFFF} 
                \cellcolor[HTML]{000000}{\color[HTML]{FFFFFF} \textbf{Adversarial-IN+IM+T}}   & wurzburg                                         & 0                                               \\ \hline
                \rowcolor[HTML]{819171} 
                \cellcolor[HTML]{000000}{\color[HTML]{FFFFFF} \textbf{Adversarial-IN+IM+T+O}} & wurzburg                                         & 0                                               \\ \hline
                \rowcolor[HTML]{FFFFFF} 
                \cellcolor[HTML]{000000}{\color[HTML]{FFFFFF} \textbf{Adversarial-IN+O}}      & wurzburg                                         & 0                                               \\ \hline
                \rowcolor[HTML]{819171} 
                \cellcolor[HTML]{000000}{\color[HTML]{FFFFFF} \textbf{Adversarial-IN+T}}      & wurzburg                                         & 0                                               \\ \hline
                \rowcolor[HTML]{FFFFFF} 
                \cellcolor[HTML]{000000}{\color[HTML]{FFFFFF} \textbf{Adversarial-IN+T+O}}    & wurzburg                                         & 0                                               \\ \hline
                \rowcolor[HTML]{819171} 
                \cellcolor[HTML]{000000}{\color[HTML]{FFFFFF} \textbf{Adversarial-O}}         & wurzburg                                         & 0.43                                            \\ \hline
                \rowcolor[HTML]{FFFFFF} 
                \cellcolor[HTML]{000000}{\color[HTML]{FFFFFF} \textbf{Adversarial-T}}         & wurzburg                                         & 0.17                                            \\ \hline
                \rowcolor[HTML]{819171} 
                \cellcolor[HTML]{000000}{\color[HTML]{FFFFFF} \textbf{Adversarial-T+O}}       & wurzburg                                         & 0.06                                            \\ \hline
                \rowcolor[HTML]{0F52BA} 
                \cellcolor[HTML]{000000}{\color[HTML]{FFFFFF} \textbf{Adversarial-IM}}        & {\color[HTML]{FFFFFF} minmax}                    & {\color[HTML]{FFFFFF} 0.01}                     \\ \hline
                \rowcolor[HTML]{FFFFFF} 
                \cellcolor[HTML]{000000}{\color[HTML]{FFFFFF} \textbf{Adversarial-IM+O}}      & minmax                                           & 0.01                                            \\ \hline
                \rowcolor[HTML]{0F52BA} 
                \cellcolor[HTML]{000000}{\color[HTML]{FFFFFF} \textbf{Adversarial-IM+T}}      & {\color[HTML]{FFFFFF} minmax}                    & {\color[HTML]{FFFFFF} 0}                        \\ \hline
                \rowcolor[HTML]{FFFFFF} 
                \cellcolor[HTML]{000000}{\color[HTML]{FFFFFF} \textbf{Adversarial-IM+T+O}}    & minmax                                           & 0.01                                            \\ \hline
                \rowcolor[HTML]{0F52BA} 
                \cellcolor[HTML]{000000}{\color[HTML]{FFFFFF} \textbf{Adversarial-IN}}        & {\color[HTML]{FFFFFF} minmax}                    & {\color[HTML]{FFFFFF} 0}                        \\ \hline
                \rowcolor[HTML]{FFFFFF} 
                \cellcolor[HTML]{000000}{\color[HTML]{FFFFFF} \textbf{Adversarial-IN+IM}}     & minmax                                           & 0                                               \\ \hline
                \rowcolor[HTML]{0F52BA} 
                \cellcolor[HTML]{000000}{\color[HTML]{FFFFFF} \textbf{Adversarial-IN+IM+O}}   & {\color[HTML]{FFFFFF} minmax}                    & {\color[HTML]{FFFFFF} 0}                        \\ \hline
                \rowcolor[HTML]{FFFFFF} 
                \cellcolor[HTML]{000000}{\color[HTML]{FFFFFF} \textbf{Adversarial-IN+IM+T}}   & minmax                                           & 0                                               \\ \hline
                \rowcolor[HTML]{0F52BA} 
                \cellcolor[HTML]{000000}{\color[HTML]{FFFFFF} \textbf{Adversarial-IN+IM+T+O}} & {\color[HTML]{FFFFFF} minmax}                    & {\color[HTML]{FFFFFF} 0}                        \\ \hline
                \rowcolor[HTML]{FFFFFF} 
                \cellcolor[HTML]{000000}{\color[HTML]{FFFFFF} \textbf{Adversarial-IN+O}}      & minmax                                           & 0                                               \\ \hline
                \rowcolor[HTML]{0F52BA} 
                \cellcolor[HTML]{000000}{\color[HTML]{FFFFFF} \textbf{Adversarial-IN+T}}      & {\color[HTML]{FFFFFF} minmax}                    & {\color[HTML]{FFFFFF} 0}                        \\ \hline
                \rowcolor[HTML]{FFFFFF} 
                \cellcolor[HTML]{000000}{\color[HTML]{FFFFFF} \textbf{Adversarial-IN+T+O}}    & minmax                                           & 0                                               \\ \hline
                \rowcolor[HTML]{0F52BA} 
                \cellcolor[HTML]{000000}{\color[HTML]{FFFFFF} \textbf{Adversarial-O}}         & {\color[HTML]{FFFFFF} minmax}                    & {\color[HTML]{FFFFFF} 0.33}                     \\ \hline
                \rowcolor[HTML]{FFFFFF} 
                \cellcolor[HTML]{000000}{\color[HTML]{FFFFFF} \textbf{Adversarial-T}}         & minmax                                           & 0.42                                            \\ \hline
                \rowcolor[HTML]{0F52BA} 
                \cellcolor[HTML]{000000}{\color[HTML]{FFFFFF} \textbf{Adversarial-T+O}}       & {\color[HTML]{FFFFFF} minmax}                    & {\color[HTML]{FFFFFF} 0.09}                     \\ \hline
                \caption{\texttt{imposters()} Experiment (Full-Text Supplied): The configuration for \texttt{imposters()}'s parameters was as follows: the \texttt{test} parameter, representing the ``text to be checked for authorship,'' was set to ``Hughes\_CypherpunkManifesto.txt;'' the \texttt{candidate.set}, which represents ``a table containing frequencies/counts for the candidate set'' was set to the collection of adversarially modified versions of ``Hughes\_CypherpunkManifesto.txt'' (see (\textbf{Figure \ref{fig:Training_and_Test_Data_List}})); the function's return value is ``a single score indicating the probability that an anonymou[s] sample analyzed was [(or was not)] written by a candidate author.''}
                \label{tab:Stylo_Imposters_Probability_Scores_Full}
            \end{longtable}

            \clearpage

            \begin{longtable}{|l|l|l|}
                \hline
                \rowcolor[HTML]{000000} 
                {\color[HTML]{FFFFFF} }                                                        & {\color[HTML]{FFFFFF} \textbf{Distance Measure}} & {\color[HTML]{FFFFFF} \textbf{Imposters Score}} \\ \hline
                \endfirsthead

                \multicolumn{3}{l}{\itshape (continued from previous page)} \\
                \endhead

                \multicolumn{3}{r}{\itshape (continued on the next page)} \\
                \endfoot

                \endlastfoot

                \rowcolor[HTML]{97551A} 
                \cellcolor[HTML]{000000}{\color[HTML]{FFFFFF} \textbf{Adversarial-IN-DI}} & {\color[HTML]{FFFFFF} delta}                     & {\color[HTML]{FFFFFF} 0.02}                     \\ \hline
                \cellcolor[HTML]{000000}{\color[HTML]{FFFFFF} \textbf{Adversarial-IN-LI}} & delta                                            & 0.52                                            \\ \hline
                \rowcolor[HTML]{893244} 
                \cellcolor[HTML]{000000}{\color[HTML]{FFFFFF} \textbf{Adversarial-IN-DI}} & {\color[HTML]{FFFFFF} argamon}                   & {\color[HTML]{FFFFFF} 0.05}                     \\ \hline
                \cellcolor[HTML]{000000}{\color[HTML]{FFFFFF} \textbf{Adversarial-IN-LI}} & argamon                                          & 0.48                                            \\ \hline
                \rowcolor[HTML]{A6A6ED} 
                \cellcolor[HTML]{000000}{\color[HTML]{FFFFFF} \textbf{Adversarial-IN-DI}} & eder                                             & 0.05                                            \\ \hline
                \cellcolor[HTML]{000000}{\color[HTML]{FFFFFF} \textbf{Adversarial-IN-LI}} & eder                                             & 0.44                                            \\ \hline
                \rowcolor[HTML]{551A97} 
                \cellcolor[HTML]{000000}{\color[HTML]{FFFFFF} \textbf{Adversarial-IN-DI}} & {\color[HTML]{FFFFFF} simple}                    & {\color[HTML]{FFFFFF} 0}                        \\ \hline
                \cellcolor[HTML]{000000}{\color[HTML]{FFFFFF} \textbf{Adversarial-IN-LI}} & simple                                           & 0.55                                            \\ \hline
                \rowcolor[HTML]{28465B} 
                \cellcolor[HTML]{000000}{\color[HTML]{FFFFFF} \textbf{Adversarial-IN-DI}} & {\color[HTML]{FFFFFF} canberra}                  & {\color[HTML]{FFFFFF} 0}                        \\ \hline
                \cellcolor[HTML]{000000}{\color[HTML]{FFFFFF} \textbf{Adversarial-IN-LI}} & canberra                                         & 0.49                                            \\ \hline
                \rowcolor[HTML]{6082B6} 
                \cellcolor[HTML]{000000}{\color[HTML]{FFFFFF} \textbf{Adversarial-IN-DI}} & manhattan                                        & 0.1                                             \\ \hline
                \cellcolor[HTML]{000000}{\color[HTML]{FFFFFF} \textbf{Adversarial-IN-LI}} & manhattan                                        & 0.48                                            \\ \hline
                \rowcolor[HTML]{FFBF00} 
                \cellcolor[HTML]{000000}{\color[HTML]{FFFFFF} \textbf{Adversarial-IN-DI}} & euclidean                                        & 0.39                                            \\ \hline
                \cellcolor[HTML]{000000}{\color[HTML]{FFFFFF} \textbf{Adversarial-IN-LI}} & euclidean                                        & 0.02                                            \\ \hline
                \rowcolor[HTML]{00674F} 
                \cellcolor[HTML]{000000}{\color[HTML]{FFFFFF} \textbf{Adversarial-IN-DI}} & {\color[HTML]{FFFFFF} cosine}                    & {\color[HTML]{FFFFFF} 0}                        \\ \hline
                \cellcolor[HTML]{000000}{\color[HTML]{FFFFFF} \textbf{Adversarial-IN-LI}} & cosine                                           & 0.56                                            \\ \hline
                \rowcolor[HTML]{819171} 
                \cellcolor[HTML]{000000}{\color[HTML]{FFFFFF} \textbf{Adversarial-IN-DI}} & wurzburg                                         & 0.02                                            \\ \hline
                \cellcolor[HTML]{000000}{\color[HTML]{FFFFFF} \textbf{Adversarial-IN-LI}} & wurzburg                                         & 0.47                                            \\ \hline
                \rowcolor[HTML]{0F52BA} 
                \cellcolor[HTML]{000000}{\color[HTML]{FFFFFF} \textbf{Adversarial-IN-DI}} & {\color[HTML]{FFFFFF} minmax}                    & {\color[HTML]{FFFFFF} 0}                        \\ \hline
                \cellcolor[HTML]{000000}{\color[HTML]{FFFFFF} \textbf{Adversarial-IN-LI}} & minmax                                           & 0.54    \\ \hline                                       
                \caption{\texttt{imposters()} Experiment (Full-Text Supplied): As a continuation of (\textbf{Figure \ref{fig:Stylo_Classify_Final_Results_Full_LI_DI}} and \textbf{Table \ref{tab:Stylo_Classify_Final_Results_Distance_Table_Full_LI_DI}}), the \texttt{imposters()} results further reinforce the subcomponent hierarchy: Doppelg\"anger Injection \( > \) Liminal Injection.}
                \label{tab:Stylo_Imposters_Probability_Scores_Full_LI_DI}
            \end{longtable}

        \subsection{\texttt{crossv()} Experiments}
        \label{appx:Crossv}

            \begin{figure}[H]
                \centering
                \subfloat{\includegraphics[width=0.41\linewidth]{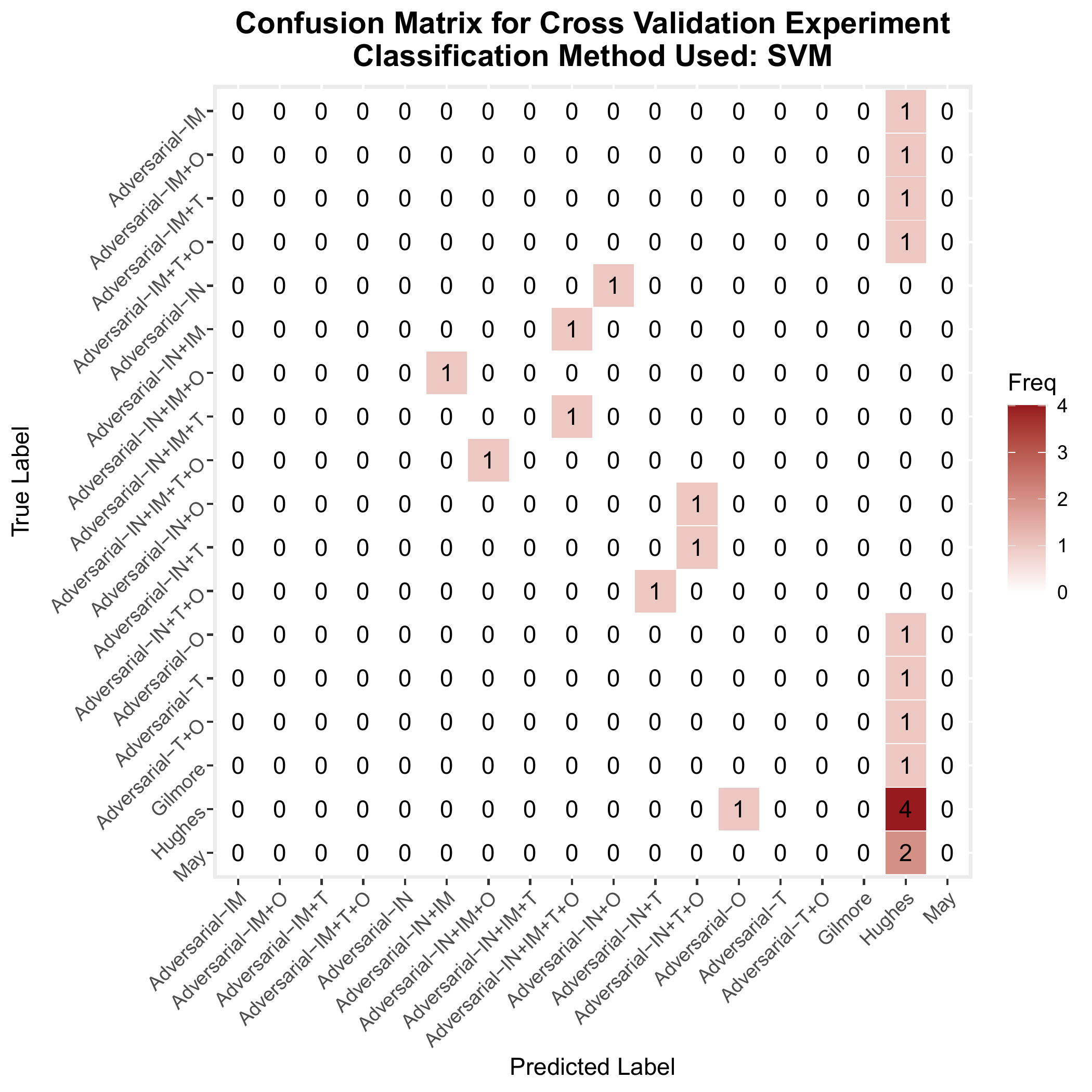}} \quad
                \subfloat{\includegraphics[width=0.41\linewidth]{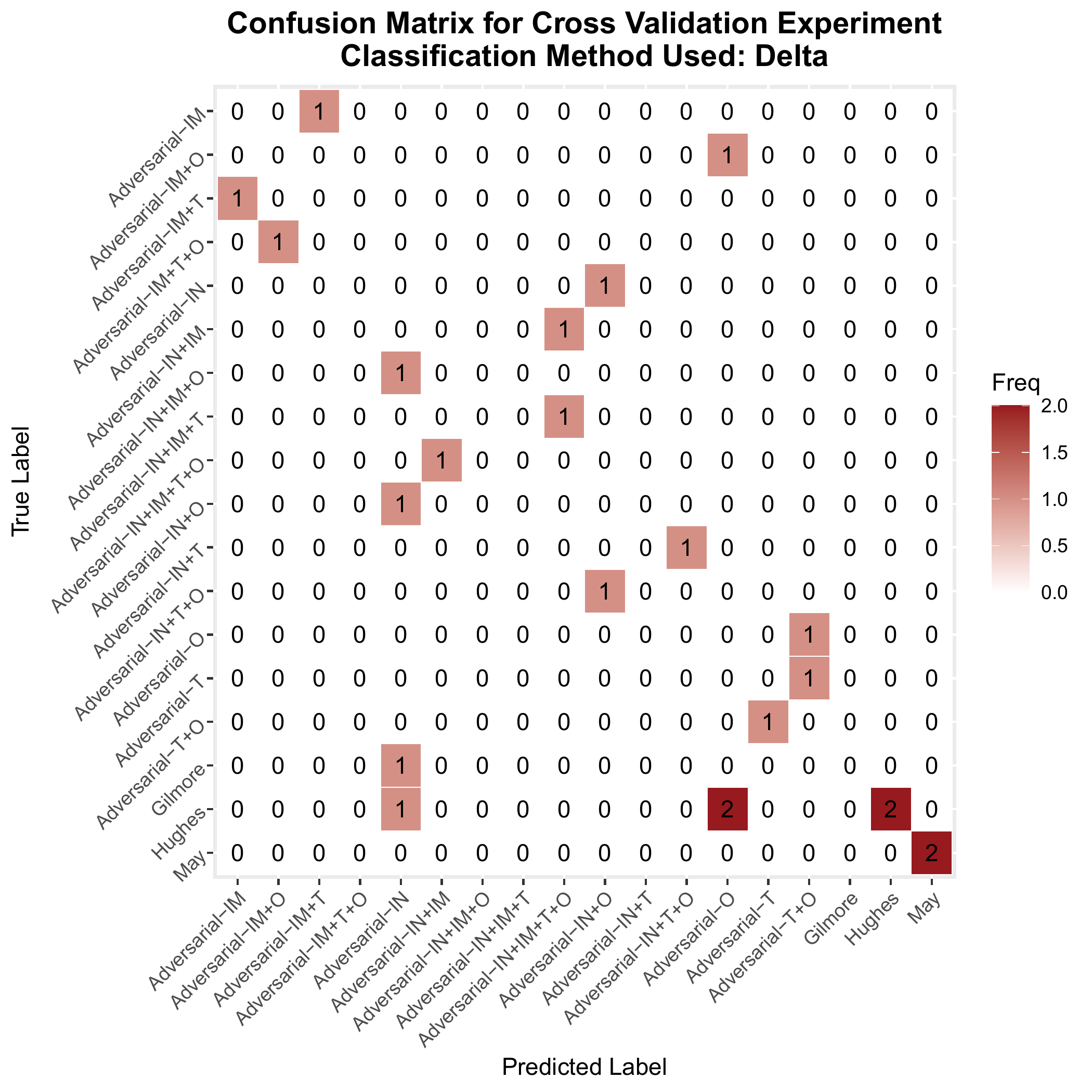}} \quad
                \subfloat{\includegraphics[width=0.41\linewidth]{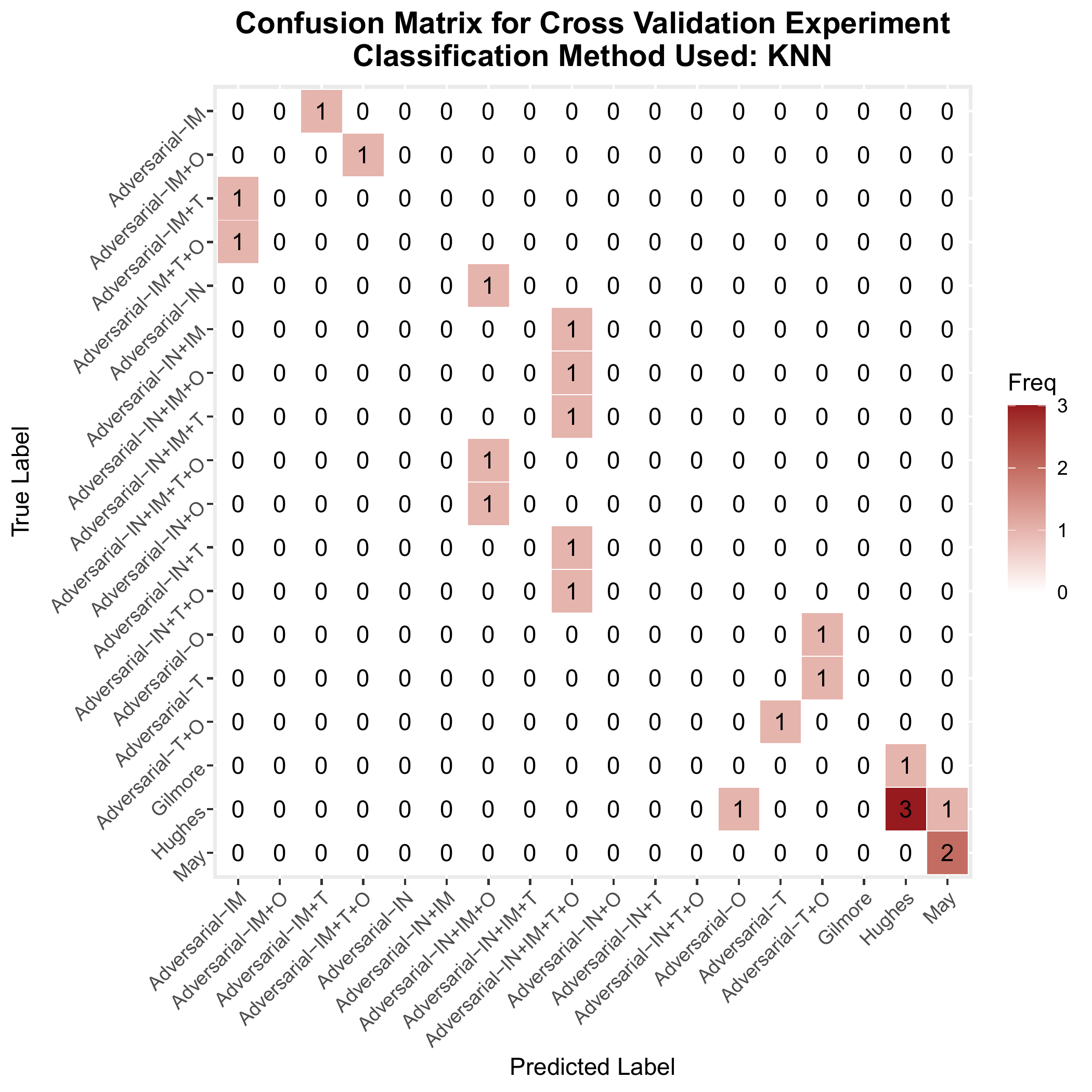}} \quad
                \subfloat{\includegraphics[width=0.41\linewidth]{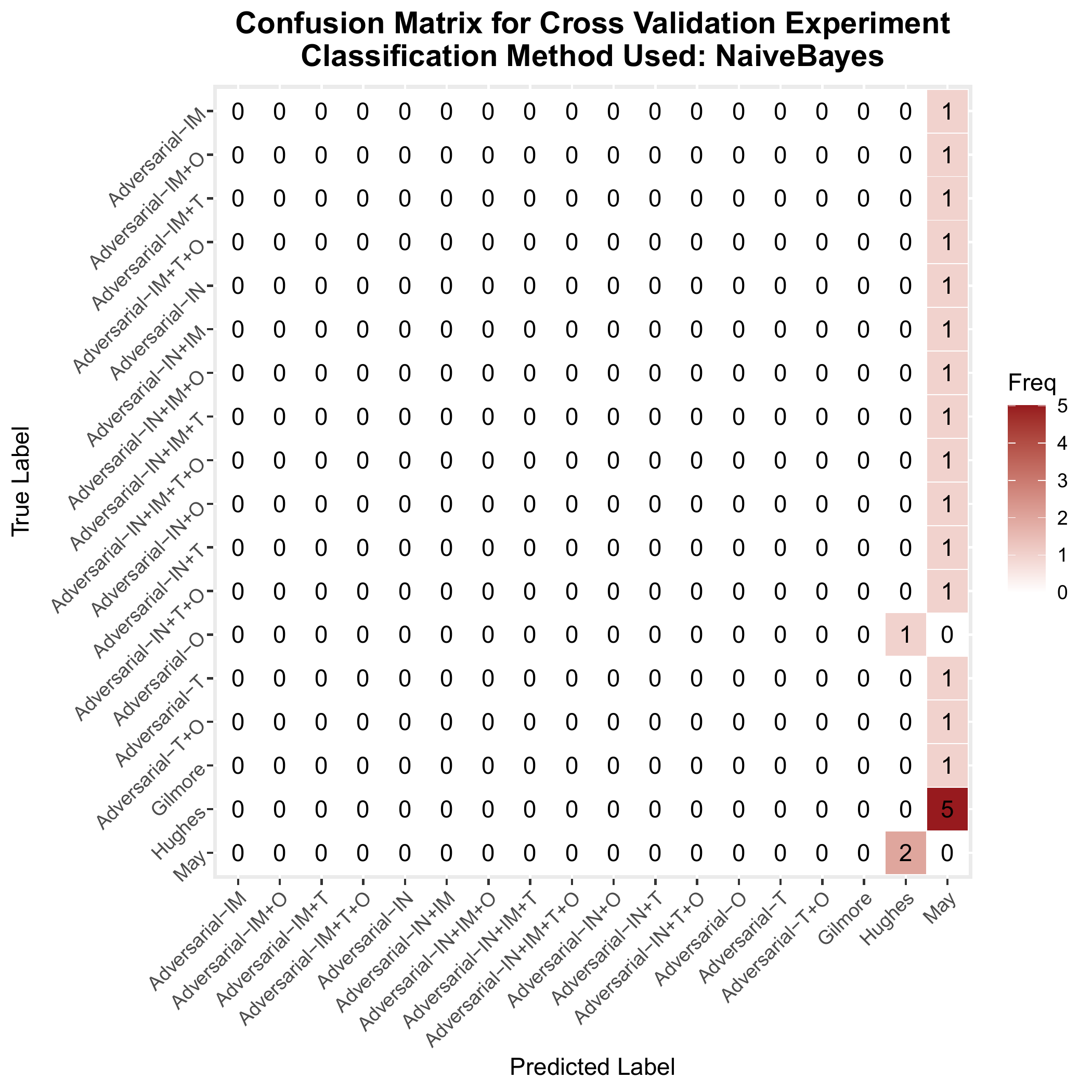}}
                \caption{\texttt{crossv()} Experiment (Full-Text Supplied): On the whole, Obfuscation seems to fare far worse than Translation when using the full texts rather than snippets of the originals. Using a majority rather than a modicum of the text implies that the actual ranking of techniques is Injection \( > \) Imitation \( > \) Translation \( > \) Obfuscation, which is informed by both the \texttt{crossv()} confusion matrices and the \texttt{imposters()} scores (see (\textbf{Table \ref{tab:Stylo_Imposters_Probability_Scores_Full}})). Irrespective of the distance measure used, the scores for Injection were almost always zero, Imitation ranged from 0.01--0.05, Translation ranged from 0.16--0.49, and Obfuscation ranged from 0.19--0.59 (the Obfuscation interval was wider than Translation's by 0.07). Scores that fall roughly between 0.39 and 0.63 are deemed suspicious, indicating that the classifier was probably uncertain; scores below 0.5 (i.e., outside the suspicious interval) signal a verification failure. Regarding the confusion matrices, the \texttt{classification.method} parameters were set to Support Vector Machine (``svm''), Burrows' Delta (``delta''), k-Nearest Neighbors (``knn''), and Na\"{\i}ve Bayes (``naivebayes''). \texttt{crossv()} itself iteratively performs a classification, with the composition of the training and test sets being shuffled with each iteration.}
                \label{fig:Stylo_Crossv_Confusion_Matrices_Full}
            \end{figure}

    \section{Key Terminology and Corresponding Definitions}
    \label{appx:Definitions}

        \epigraph{\textcolor{adversarial}{It'ѕ аn еуе fоr аn еуе / Тһеn а һеаd fоr tһе еуе / Тһеn а lіfе fоr tһе еуе / Тһеn tһе vіllаgе fоr tһе еуе / Тһеn а сіtу fоr tһе еуе / Тһеn а соuntrу fоr tһе еуе / Тһе wһоlе\ldots wоrld burnѕ fоr tһе еуе}}{\textit{{\scriptsize Ѕ\hspace{0pt}аm С\hspace{0pt}аrtеr, R\hspace{0pt}оb D\hspace{0pt}аmіаnі, Ѕ\hspace{0pt}іmоn Dеlаnеу, М\hspace{0pt}аtt D\hspace{0pt}оnnеllу, Т\hspace{0pt}оm D\hspace{0pt}оуlе}}}

        \begin{itemize}
            \item[\ding{118}] \textbf{Support Vector Machine (SVM)}: a binary-classification algorithm that finds the optimal separating hyperplane by maximizing the margin between the nearest points (support vectors); it handles overlapping classes with a soft-margin that down-weights misclassified points, and non-linear separations via kernel-induced high-dimensional projections, solving the problem as a quadratic optimization (\textit{Meyer} \cite{Meyer2025}).
            \item[\ding{118}] \textbf{Burrows' Delta}: a supervised classification technique that calculates a distance matrix between texts, then assigns each test sample to the class of its nearest training neighbor based on those distances \cite{DBC}.
            \item[\ding{118}] \textbf{k-Nearest Neighbors (k-NN)}: a non-parametric, supervised learning met\-hod for classification and regression that assigns a label (by majority/plurality vote) or predicts a value for a new instance based on the \( k \) most similar (nearest) examples, using distance as a measure of similarity (\textit{Bhandari} \cite{Bhandari2025}).
            \item[\ding{118}] \textbf{Na\"{\i}ve Bayes}: a family of probabilistic classifiers that apply Bayes' theorem with the simplifying assumption that features are conditionally independent, estimating class posteriors as the product of the class prior and individual feature likelihoods and predicting the class with the highest posterior probability (\textit{Majka} \cite{Majka2024}).
            \item[\ding{118}] \textbf{Principal Components Analysis (PCA)}: a dimensionality-reduction technique that linearly transforms correlated continuous variables into orthogonal ``principal components'' ordered by decreasing explained variance, allowing a lower-dimensional representation that retains most of the original information \cite{PCA2026}.
            \item[\ding{118}] \textbf{Bootstrap Consensus Tree}: a method that aggregates many dendrograms (generated from multiple ``snapshots'' of data, such as varying numbers of most-frequent words) into a single consensus diagram, highlighting groupings that repeatedly appear across the bootstrapped trees as robust patterns \cite{BCN}.
            \item[\ding{118}] \textbf{Cluster Analysis}: an exploratory data-mining technique that partitions a set of objects into clusters so that items within each cluster are more similar to each other than to items in other clusters (\textit{Mouselimis} \cite{Mouselimis2025}).
            \item[\ding{118}] \textbf{Text Stylometry}: the quantitative analysis of writing style---employing computational pipelines of preprocessing, feature extraction, and statistical modeling---to identify patterns such as authorship, age, gender, or other metadata across large text collections. This involves extracting stylistic features (e.g., function-word frequencies, letter- or word-level \( n \)-grams, part-of-speech distributions) and measuring vocabulary richness or other overall stylistic metrics to infer demographic attributes and other contextual information (\textit{Eder et al.} \cite{Eder2016}; \textit{Brozovsky} \cite{Brozovsky2024}).
            \item[\ding{118}] \textbf{Adversarial Stylometry}: the study of deliberately modifying or imitating writing style---through Obfuscation, Imitation, Translation, or Injection---to defeat authorship-attribution methods, enabling writers to protect their privacy and maintain anonymity by rendering stylometric classifiers no more accurate than random guessing (\textit{Brennan et al.} \cite{Brennan2012}).
            \item[\ding{118}] \textbf{Text Steganography}: the technique of embedding hidden information with\-in natural language text so that the presence of the secret data remains imperceptible to anyone except the intended recipient (\textit{Ahvanooey et al.} \cite{Ahvanooey2019}; \textit{Thereallo} \cite{Thereallo2026}).
                \begin{itemize}
                    \item[\eye] \textbf{Zero-width Unicode Characters}: a sampling of Unicode characters that produce no visible symbol or width includes the \textit{Zero-Width-Non-Joiner} \texttt{[U+200C]}, \textit{Left-To-Right Mark} \texttt{[U+200E]}, \textit{Right-To-Left Mark} \texttt{[U+200F]}, \textit{Zero-Width-Joiner} \texttt{[U+200D]}, \textit{Zero-Width-Space} \texttt{[U+200B]}, and \textit{Zero-Width-Non-Break} \texttt{[U+FEFF]}. A more exhaustive list of these characters and how they can be utilized for steganography can be found in \textit{Ahvanooey et al.} \cite{Ahvanooey2019}. Watermarking with special characters \cite{Rumi2025} such as the \textit{Narrow No-Break Space} \texttt{[U+202F]} (not zero-width) is relevant here because it exploits a familiar paradigm: a character looks identical to another character in most word processors and browsers, making it virtually impossible to distinguish visually.
                    \item[\eye] \textbf{Homoglyph Substitution}: the technique of replacing characters in a text with visually similar characters from different scripts or Unicode code points to conceal or alter the content while preserving its appearance. It can be used to encode and embed secret data within text, though the approach is most practical when applied to Latin-based cover texts. For example, try searching for the word ``death'' in the digital version of this document using the built-in search function. You should detect only one instance, but another occurrence appears in the text. The first occurrence has been obscured using the technique (\textit{Rizzo et al.} \cite{Rizzo2016}).
                \end{itemize}
        \end{itemize}

    \section{Formulation of Distance Metrics and Their Mathematical Expressions}
    \label{appx:Formulas}

        \epigraph{\textcolor{adversarial}{My eyes were dimmed and filled with thousands of sinusoids\ldots}}{\textit{{\scriptsize We \\ Yevgeny Zamyatin}}}

        In stylometry, a \textbf{distance measure} is a function that takes a matrix of word-frequency (or other feature) counts for a set of texts, optionally transforms the data, and returns a symmetric square matrix whose entries quantify the dissimilarity between every pair of texts. The matrix's diagonal is zero (identical texts have no distance), and larger values indicate greater stylistic divergence (\textit{Eder} \cite{Eder2015}).

        Below, we delineate the fundamental formulas underlying the distance metrics employed in our study. Regarding notation, we use uppercase symbols to represent vectors (or points) and a lowercase symbol with a subscript to refer to a specific component of a vector. Accordingly, \( A \), \( B \), and \( C \) denote vectors; the notation \( a_{i} \) designates the \( i \)-th entry of vector \( A \); and \( m \) indicates the dimension (length) of the vectors. For each position \( i \), \( \sigma_{i} \) and \( \mu_{i} \) denote the standard deviation and mean, respectively, of the values across all vectors under consideration. The symbols \( \min \) and \( \max \) refer to the smallest and largest values, respectively (\textit{Stanik\={u}nas et al.} \cite{Donatas2017}; \textit{Kocher and Savoy} \cite{Kocher2017}).

        \begin{itemize}
            \item[\ding{118}] \textbf{Burrows' Delta (\textcolor[HTML]{97551A}{delta})}: \newline \( \frac{1}{m}\sum_{i=1}^{m}\left | \left ( \frac{a_{i} - \mu_{i}}{\sigma_{i}} \right ) - \left ( \frac{b_{i} - \mu_{i}}{\sigma_{i}} \right ) \right | \) \newline
            \item[\ding{118}] \textbf{Argamon's Linear Delta (\textcolor[HTML]{893244}{argamon})}: \newline \( \frac{1}{m}\sum_{i=1}^{m}\sqrt{\left | \frac{(a_{i} - b_{i})^{2}}{\sigma_{i}^{2}}\right |} \) \newline
            \item[\ding{118}] \textbf{Eder's Delta (\textcolor[HTML]{a6a6ed}{eder})}: \newline \( \frac{1}{m}\sum_{i=1}^{m}\left ( \left | \frac{a_{i} - b_{i}}{\sigma_{i}}\right | \cdot \frac{\left ( m - m_{i} + 1 \right ) }{m}\right ) \) \newline
            \item[\ding{118}] \textbf{Eder's Simple Distance (\textcolor[HTML]{551a97}{simple})}: \newline \( \sum_{i=1}^{m}\left | \sqrt{a_{i}} - \sqrt{b_{i}} \right | \) \newline
            \item[\ding{118}] \textbf{Canberra Distance (\textcolor[HTML]{28465B}{canberra})}: \newline \( \sum_{i=1}^{m}\frac{\left | a_{i} - b_{i} \right |}{\left | a_{i} \right | + \left | b_{i} \right |} \) \newline
            \item[\ding{118}] \textbf{Manhattan Distance (\textcolor[HTML]{6082B6}{manhattan})}: \newline \( \sum_{i=1}^{m}\left | a_{i} - b_{i} \right | \) \newline
            \item[\ding{118}] \textbf{Euclidean Distance (\textcolor[HTML]{ffbf00}{euclidean})}: \newline \( \sqrt{\sum_{i=1}^{m} \left ( a_{i} - b_{i} \right )^{2} } \) \newline
            \item[\ding{118}] \textbf{Cosine Distance (\textcolor[HTML]{00674F}{cosine})}: \newline \( \frac{\sum_{i=1}^{m} a_{i} \cdot b_{i}}{\sqrt{\sum_{i=1}^{m} a_{i}^{2}} \cdot \sqrt{\sum_{i=1}^{m} b_{i}^{2}}} \) \newline
            \item[\ding{118}] \textbf{Cosine Delta Distance (\textcolor[HTML]{819171}{wurzburg}) \cite{CDD,Smith2011,Evert2017}}: \newline \( \frac{\sum_{i=1}^{m} \frac{a_{i} - \mu_{i}}{\sigma_{i}} \cdot \frac{b_{i} - \mu_{i}}{\sigma_{i}}}{\sqrt{\left ( \sum_{i=1}^{m} \frac{a_{i} - \mu_{i}}{\sigma_{i}} \right )^{2}} \cdot \sqrt{\left ( \sum_{i=1}^{m} \frac{b_{i} - \mu_{i}}{\sigma_{i}} \right )^{2}}} \) \newline
            \item[\ding{118}] \textbf{Min-Max Distance (\textcolor[HTML]{0F52BA}{minmax}) \cite{MMD,Kestemont2016}}: \newline \( \frac{\sum_{i=1}^{m} \min\left ( a_{i},b_{i} \right )}{\sum_{i=1}^{m} \max\left ( a_{i},b_{i} \right )} \)
        \end{itemize}

\end{subappendices}

\end{document}